\documentclass[%
 aip,pop,
 amsmath,amssymb,
 reprint,%
]{revtex4-1}

\usepackage{graphicx}
\usepackage{dcolumn}
\usepackage{bm}

\usepackage[utf8]{inputenc}
\usepackage[T1]{fontenc}
\usepackage{mathptmx}

\usepackage{todonotes} 

\newcommand{\mpon}{\textcolor{black}}

\newcommand{\ab}{\textcolor{black}}
\newcommand{\mpo}{\textcolor{black}}
\newcommand{\pjm}{\textcolor{black}}

\newcommand{\abrev}{\textcolor{black}}
\newcommand{\mwein}{\textcolor{black}}

\usepackage{soul} 

\usepackage{epstopdf}

\newcommand{\thbn}{\theta_{\rm Bn}}

\newcommand{\ompe}{\omega_\mathrm{pe}}
\newcommand{\ompi}{\omega_\mathrm{pi}}
\newcommand{\ompb}{\omega_\mathrm{pb}}
\newcommand{\omce}{\Omega_\mathrm{e}}
\newcommand{\omci}{\Omega_\mathrm{i}}

\newcommand{\ms}{M_\mathrm{s}}

\newcommand{\ma}{M_\mathrm{A}}
\newcommand{\mi}{m_\mathrm{i}}
\newcommand{\me}{m_\mathrm{e}}
\newcommand{\lse}{\lambda_\mathrm{se}}
\newcommand{\lsi}{\lambda_\mathrm{si}}
\newcommand{\vsh}{v_\mathrm{sh}}
\newcommand{\vdr}{v_\mathrm{dr}}
\newcommand{\omres}{\varpi_{\mathrm{res}}}
\newcommand{\II}{\mathrm{i}}

\def\ee{\end{equation}}
\def\be{\begin{equation}}

\def\apjl{ApJ Lett.}%
\def\apj{ApJ}%
\def\jgr{JGR}%
\def\grl{Geo. Res. Lett.}%

\begin{document}

\preprint{AIP/123-QED}

\title{The electron foreshock at high-Mach-number \\ nonrelativistic oblique shocks}

\author{Artem Bohdan}
\email{artem.bohdan@desy.de}
\affiliation{Deutsches Elektronen-Synchrotron DESY, Platanenallee 6, 15738 Zeuthen, Germany}

\author{Martin~S.~Weidl}
\affiliation{Max-Planck-Institut für Plasmaphysik, Boltzmannstr. 2, DE-85748 Garching, Germany}

\author{Paul~J.~Morris}
\affiliation{Deutsches Elektronen-Synchrotron DESY, Platanenallee 6, 15738 Zeuthen, Germany}

\author{Martin Pohl}
\affiliation{Deutsches Elektronen-Synchrotron DESY, Platanenallee 6, 15738 Zeuthen, Germany}
\affiliation{Institute of Physics and Astronomy, University of Potsdam, DE-14476 Potsdam, Germany}

\date{\today}

\begin{abstract}
\ab{\mpon{In the Universe matter outside of stars and} compact objects is mostly composed of collisionless plasma. The interaction of a supersonic plasma flow with an obstacle results in collisionless shocks \mpon{that are} often associated with intense nonthermal radiation and the production of cosmic ray particles.} Motivated by simulations of non-relativistic high-Mach-number shocks in supernova remnants, we investigate the instabilities excited by relativistic electron beams in the extended foreshock of oblique shocks. The phase-space distributions in the \abrev{inner and outer foreshock regions are derived with a Particle-in-Cell simulation of the shock and} used as initial conditions for simulations with periodic boundary conditions to study their relaxation towards equilibrium. We find that the observed electron-beam instabilities agree very well with the predictions of a linear dispersion analysis: the electrostatic electron-acoustic instability dominates in the \abrev{outer region} of the foreshock, while the denser electron beams in the \abrev{inner foreshock} drive the gyroresonant oblique-whistler instability.
\end{abstract}

\maketitle

%

\section{Introduction} \label{sec:intro}

The astrophysical environment is often associated with collisionless plasma, in which the collisional mean free path is much larger \mpo{than} the characteristic system length. The physics of such a plasma is defined by collective plasma behaviour and wave-particle interactions. Therefore, when supersonic plasma collides with an obstacle, a so-called collisionless shock is formed whose physics and energy dissipation processes are also governed by collective particle behaviour. \mpo{Intense nonthermal radiation associated with the cosmic ray (CR) particles produced at collisionless shocks has been observed from} many astrophysical system, such as supernova remnants (SNRs), active galactic nuclei, galaxy clusters, etc. Indeed, in the late 70s it was established that CRs can be efficiently accelerated via diffusive shock acceleration (DSA)\cite{1977ICRC...11..132A,1977DoSSR.234.1306K,1983RPPh...46..973D,Bell.1978a,1978ApJ...221L..29B}, also known as first-order Fermi acceleration, in which particles gain their energies in repetitive interactions with the shock front.
Non-relativistic collisionless shocks at SNRs are great objects to be studied in this context. Firstly, SNR shocks are efficient CR accelerators, and the current paradigm of CR origin states that the \mpo{bulk} of galactic CRs \pjm{are} produced by SNRs. Secondly, SNRs are close enough that the shock properties and the CR spectra can be deduced using observations of nonthermal radiation in radio-, x- and $\gamma$ rays. Although ions can generally reach higher energies, understanding the role of electrons is crucial to \mpo{correctly} interpret observations of nonthermal radiation at SNR shocks.

SNR shocks propagate with nonrelativistic velocities \cite{2009ApJ...699L.139W} and are characterized by high sonic and Alfv\'enic Mach numbers, $M_s,M_A \approx 20-1000$. \abrev{This regime is traditionally called the high Mach number regime in contrast to low Mach number heliospheric shocks (e.g., the Erath's bow shock) for which Mach numbers range on average from 2 to 10.} 

Once the shock velocity and the Mach numbers are established, the shock parameter \mpo{that most} affects the shock physics and the electron dynamic is the shock obliquity angle\cite{2009A&ARv..17..409T,2014ApJ...783...91C}, $\thbn$, which is defined as the angle between the upstream magnetic field and the shock normal vector. 
\abrev{Perpendicular shocks ($\thbn \approx 90^\circ$) have already been thoroughly studied over the last decade, therefore electron acceleration and heating processes are well understood in this regime.} Electrons can be accelerated via shock-surfing acceleration \cite{Hoshino2002,Matsumoto2012,Bohdan2017,Bohdan2019a}, magnetic reconnection \cite{Matsumoto2015,Bohdan2020a}, and stochastic Fermi-acceleration \citep{Bohdan2017}. The shock dynamic becomes more complicated if it satisfies the subluminal condition, \mpo{$\vsh < c / \tan{\thbn}$, where $\vsh$ is the shock speed} and $c$ is the speed of light. In this case energetic particles can escape the shock traveling far upstream. The shock transition becomes much \mpo{thicker}, and the shock foot is replaced by a broad, turbulent foreshock that extends far into the upstream flow. \abrev{The foreshock is defined as a region in front of the shock where the shock reflected particles are present.} Depending on the value of $\thbn$, the foreshock physics can be defined by ions ($\thbn\lesssim50^\circ$) \cite{2014ApJ...783...91C}, electrons ($50^\circ\lesssim\thbn\lesssim{70}^\circ$) \cite{Amano2007,Xu2020}, or the shock-emitted waves ($\thbn\approx 75^\circ$) \cite{2021ApJ...921L..14K}. \abrev{This suggests that the shock structure (including the foreshock) in all these cases is different, therefore, it is expected that particle acceleration processes are also different and require separate studies.}

Here we study the role of beam instabilities in the electron foreshock of oblique high Mach number shocks ($\thbn = 60^\circ$, $\ma = 30$). \abrev{Since electrons play an important role, we use the fully-kinetic treatment for plasma simulations, namely particle-in-cell (PIC) simulations, which consider all particle species as individual particles moving in the self-generated electromagnetic field. The PIC technique allows us to obtain all of the necessary information about particles (both ions and electrons) and electromagnetic fields at any given point in space and time and, therefore, to describe all details of the shock structure and acceleration processes. Early 1D PIC simulations \cite{Amano2007,Xu2020} demonstrated that} at these shocks the escaping (also called reflected) electrons streaming along the oblique magnetic field are capable of exciting various electrostatic and electromagnetic instabilities. \abrev{This regime is particularly interesting due to very efficient electron acceleration found in oblique shocks, namely stochastic shock-drift acceleration \citep{Matsumoto2017,Amano2020,Amano2022}. However the foreshock structure and the long-term evolution are poorly understood in shocks with $\thbn = 60^\circ$ and we use 2D PIC simulation to address these issues.}

\abrev{Studies of the Earth's bow shock demonstrate a variety of instabilities and waves captured at the foreshock and give us better understanding of the foreshock structure in high Mach number regime.} Relativistic electrons with up to 300~keV have recently been observed in the ion-foreshock of the Earth's bow shock\citep{Wilson16}, implying that foreshock disturbances may play a larger role in electron acceleration than previously believed. Electrostatic instabilities close to the boundary of the electron foreshock are driven by a field-aligned beam, which consists of electrons reflected at the quasiperpendicular bow shock via the Wu--Sonnerup mechanism \citep{Sonnerup69,Wu84}. Further into the electron foreshock, the phase-space distribution of electrons becomes more isotropic and only exhibits weak tails instead of an easily distinguished beam \citep{Fitzenreiter84}. The shock-reflected electrons are expected to drive Langmuir waves through a bump-on-tail instability or electron-acoustic waves through the electron-acoustic instability \citep{Thomsen83,Dum90}.

Closer to the shock, electromagnetic instabilities become much more important. \citet{Hellinger07} and \citet{Lembege09} showed in hybrid and PIC simulations that oblique whistler waves can be excited in the foot of a supercritical perpendicular shock. If these fluctuations scatter reflected ions rapidly enough, they can prevent shock reformation at intermediate Alfv{\'e}nic Mach numbers ($3<M_A<5$), but because of their obliquity these whistlers cannot be resolved in one-dimensional shock simulations.

Of course, oblique whistlers can also grow without coupling strongly to ions. The electron-firehose instability \citep{Li00,Camporeale08}, for instance, is excited if for a bi-Maxwellian electron distribution the parallel temperature exceeds the perpendicular temperature by more than a critical factor. Both propagating  and stationary (zero real frequency) electromagnetic modes can grow and eventually transform into oblique whistler waves that scatter electrons and reduce their anisotropy \citep{Hellinger14}. As an instability driven by temperature anisotropy rather than by a distinct beam of reflected particles, the electron-firehose instability has received more attention in the context of the solar wind than in the bow-shock community.

Studying beam instabilities that may be important in the solar-wind context, several groups have analyzed PIC simulations of the electrostatic two-beam instability, the electromagnetic oblique-whistler instability, and the anisotropy-driven whistler heat-flux instability \citep{Fu14,Micera20}. These simulations started with initial conditions that correspond to solar-wind parameters, often with a larger beam anisotropy than one would expect for a shock-reflected beam. For the results presented here, our initial phase space is modelled after the electron distribution in the PIC simulation of a high-Mach-number oblique shock.

The reference shock simulation and the derivation of the particle distributions are reported in Section~\ref{sec:shock}, followed by Section~\ref{sec:all-instab} with a detailed analysis of the electron instabilities that are excited as these distributions relax towards thermal equilibrium. We compare analytic and numerical results for the non-relativistic linear regime with periodic-boundary-condition simulations (further called PBCS or PBC simulations) that include all relativistic effects. Section~\ref{sec:foreshocks-length} is dedicated to an investigation of the steady-state solution for oblique shocks. In Section~\ref{sec:summary}, we summarise our findings and estimate what these results imply for electron acceleration in high-Mach-number shocks.

\section{Shock simulation} \label{sec:shock}

\subsection{Shock simulation setup} \label{sec:shock-setup}

\begin{figure*}[t]
\centering
\includegraphics[width=0.33\linewidth]{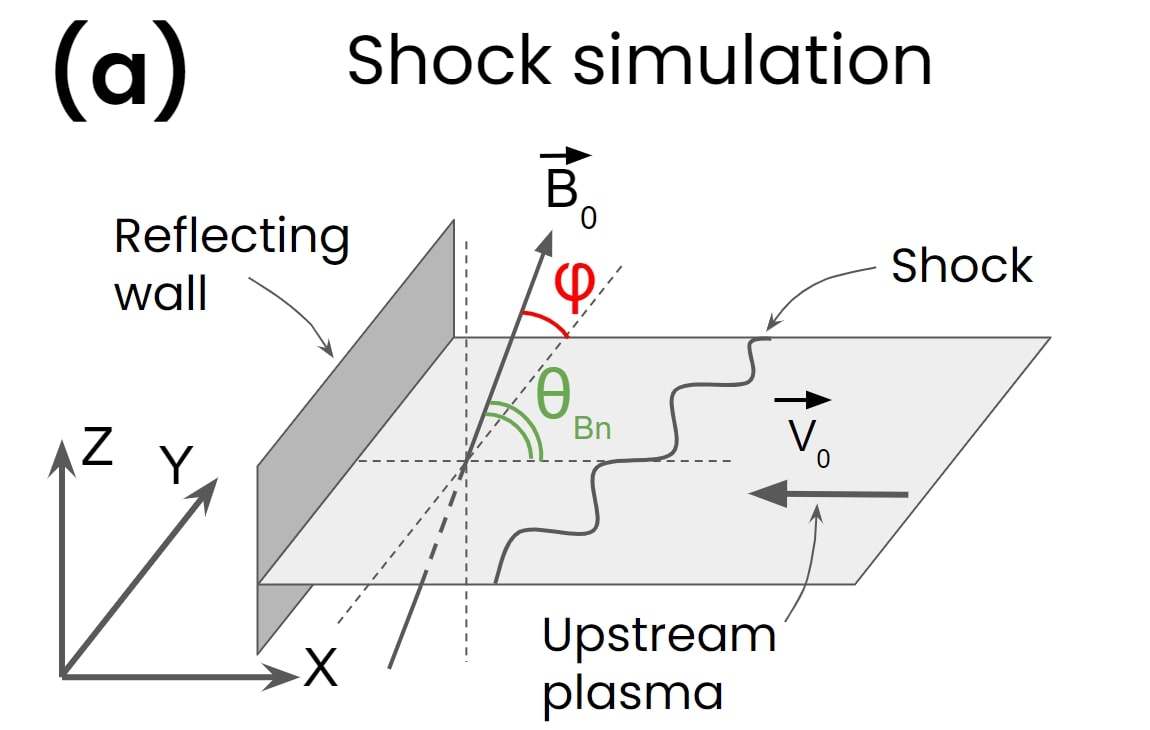}
\includegraphics[width=0.33\linewidth]{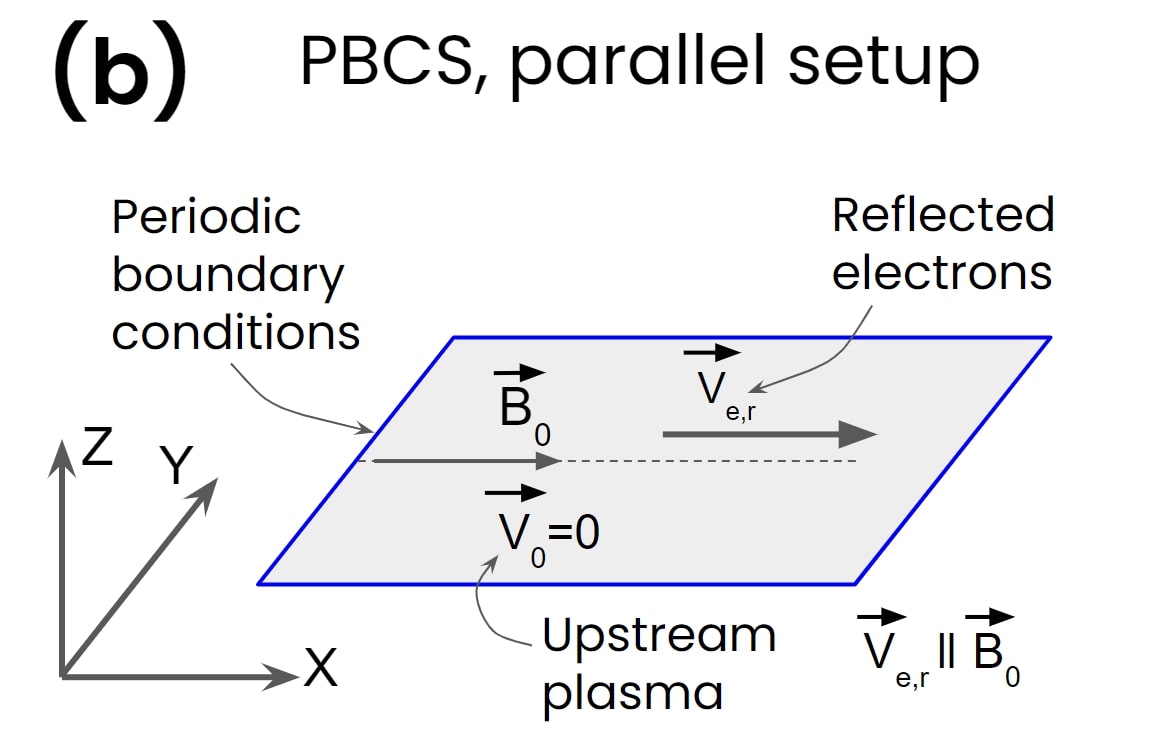}
\includegraphics[width=0.33\linewidth]{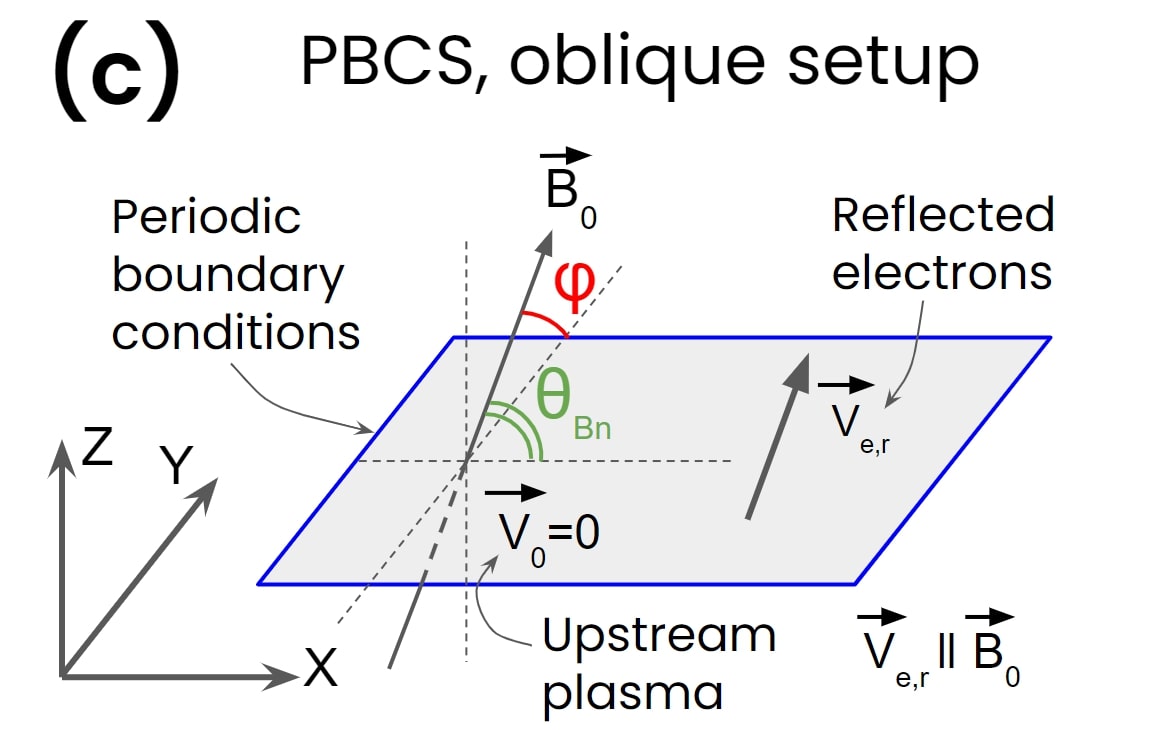}
\caption{Sketches of the geometry of the shock simulation discussed in Section~2 (a) and the PBC simulations with parallel (b) and oblique (c) magnetic field configurations presented in Section~3.}
\label{figSetup}
\end{figure*}

First, we analyse a numerical simulation of an oblique collisionless shock to obtain realistic phase-space distributions \mpo{of reflected electrons}, from which we can later derive growth rates for electron-beam driven instabilities. We use an adapted version of the fully kinetic PIC code TRISTAN \cite{Buneman1993} in a 2D3V configuration, with an MPI-based parallelization \cite{2008ApJ...684.1174N}, evolving the electric and magnetic fields on a two-dimensional Yee grid and the three momentum components of each quasiparticle with a relativistic Vay pusher \cite{Vay2008}.

\abrev{The shock simulation is performed using the reflecting wall setup which is presented in Fig.~\ref{figSetup}a.} The upstream plasma, composed of electrons and ions with density $n_0 = 40$ particles per cell per species, is initialised uniformly with a bulk velocity in the simulation frame of $\vec v_{\mathrm{up}}/c = - 0.20 \hat x$ and \mpo{the same temperature for electrons and ions,} $k_B T_e = k_B T_i = 9.86\cdot 10^{-4}~m_e c^2$, where $k_B$ is the Boltzmann constant, $m_e$ is the electron mass \abrev{and $c$ is the speed of light.} \mpo{This defines the sound speed as} $c_{\rm s}=\sqrt{\abrev{2} \Gamma k_BT_{\rm i}/\mi} = \abrev{0.0081}c$, where $\Gamma = 5/3$ is the adiabatic index and $m_i$ is the ion mass. The plasma streams left towards the $x=0$ boundary, which acts as a reflecting conducting wall (i.e.~$E_y|_{x=0}\equiv0$) \cite{1985PhRvL..54.1872Q,1989JGR....94.8783B}, \mpo{whereupon a shock forms that propagates to the right}. 
A uniform magnetic field, $\vec B_0 = B_0 (\cos \thbn, \sin \thbn \cos \varphi,\sin  \thbn \sin \varphi) = B_0 (0.5, 0, \sqrt{3}/2)$, is applied across the simulation domain with an out-of-plane component ($\thbn=60^\circ$, $\varphi=90^{\circ}$, see Fig.~\ref{figSetup}a). \abrev{As the magnetic field is assumed to be frozen in the moving plasma, a motional electric field $\vec E_0 = - \vec v_{up} \times \vec B_0$ is also initialized in the upstream region.} To smooth initial transients arising from the large $\nabla \times \vec{E}$ at $x=0$ and the corresponding $\partial \vec{B} / \partial t$, the fields are tapered to zero from the upstream value over the 50 grid cells closest to the reflecting wall \cite{Wieland2016}. Furthermore, \mpo{the now finite $\nabla \times \vec{B}$ in the tapering zone is compensated by a drift current carried by the} ions, which is removed upon reflection at the simulation boundary.

   \begin{table}[!t]
         \label{tab:simparam}
         \centering
         \setlength{\tabcolsep}{8pt}
\begin{tabular}{ c c c c c c c c } 
\hline
 \noalign{\smallskip}
 $\vsh/c$ & $\ma$ & $\ms$ & $\thbn$ & $\varphi$ & $\beta$ &$\mi/\me$ &  $t_\mathrm{sim}\omci $   \\ 
 \noalign{\smallskip}
 \hline
 \noalign{\smallskip}
  0.263 &30 & 32.5 & 60 & 90 & 1 & 50 & 50.4  \\ 
\noalign{\smallskip}
\hline
\end{tabular}
\caption{\abrev{Parameters of the shock simulation. Listed are: the shock velocity, $\vsh$, the Alfv\'enic Mach number, $\ma$, the sonic Mach number, $\ms$, the orientation angles of the uniform magnetic field with respect to the x- and y-axis, $\thbn$ and $\varphi$, the plasma beta, $\beta$, the ion-to-electron mass ratio, $\mi/\me$, the total simulations time, $t_{sim}\omci$. }} 
\end{table}

Our simulation assumes an overdense plasma typical of the interstellar medium, $\omce = |e|B_0/\me = 0.06~\ompe$, where $\ompe$ denotes the electron plasma frequency, $\omce$ represents the electron gyrofrequency and $e$ is the electron charge. With a mass ratio $m_i/m_e=50$, the Alfv{\'e}n velocity is $v_{\rm A}=B_{\rm 0}/\sqrt{\mu_{\rm 0} (N_e\me+N_i\mi)}$, where $\mu_{\rm 0}$ is the vacuum permeability and $N_i=N_e=n_0$ are the ion and the electron number densities. The total plasma beta denoting the thermal-to-magnetic energy density ratio \mpo{in the upstream region} is $\beta =1$. The simulation domain is resolved with eight grid points per electron inertial length, $\lse=c/\omega_p=8\Delta$ ($\Delta$ is the simulation grid size), and measures $288~\lse$ along the $y$ direction and initially $625~\lse$ in the $x$ direction, with the ion inertial length given by $\lsi = \sqrt{\mi/\me}~ \lse$. During the simulation, the domain length along $x$ increases, as the right wall, at which new upstream plasma is continually injected, moves outwards. This is necessary to ensure that all of the electrons that get reflected back upstream at the shock front remain in the simulation box.

By using a timestep of $\delta t=1/16\,\ompe^{-1}$, we ensure that the relevant frequencies are adequately resolved. The total run time of the simulation can be expressed as a multiple of the ion gyrofrequency, $t_\mathrm{sim} = 50.4 \omci^{-1}$, where $\omci = |e|B_0/\mi$.

After a few ion gyro times, the magnetic field and the plasma in the region downstream of the rightwards-moving density gradient have completely isotropised and reach a compression ratio of close to $n_{\mathrm{dn}}/n_0 \sim 4.0$. A quasi-stationary shock propagates along the positive $x$ axis with velocity $\vec \vsh^\ast/c = 0.06~\hat x$ in the simulation frame, corresponding to a shock velocity $\vec \vsh/c = 0.263$ as measured in the upstream frame with an Alfv{\'e}nic Mach number of $M_A = \vsh/v_A = 30$ and a sonic Mach number $M_S = \vsh/c_s = \abrev{32.5}$.

\subsection{Shock structure} \label{sec:shock-result}

\begin{figure*}[!t]
    \centering
    \includegraphics[width=0.99\linewidth]{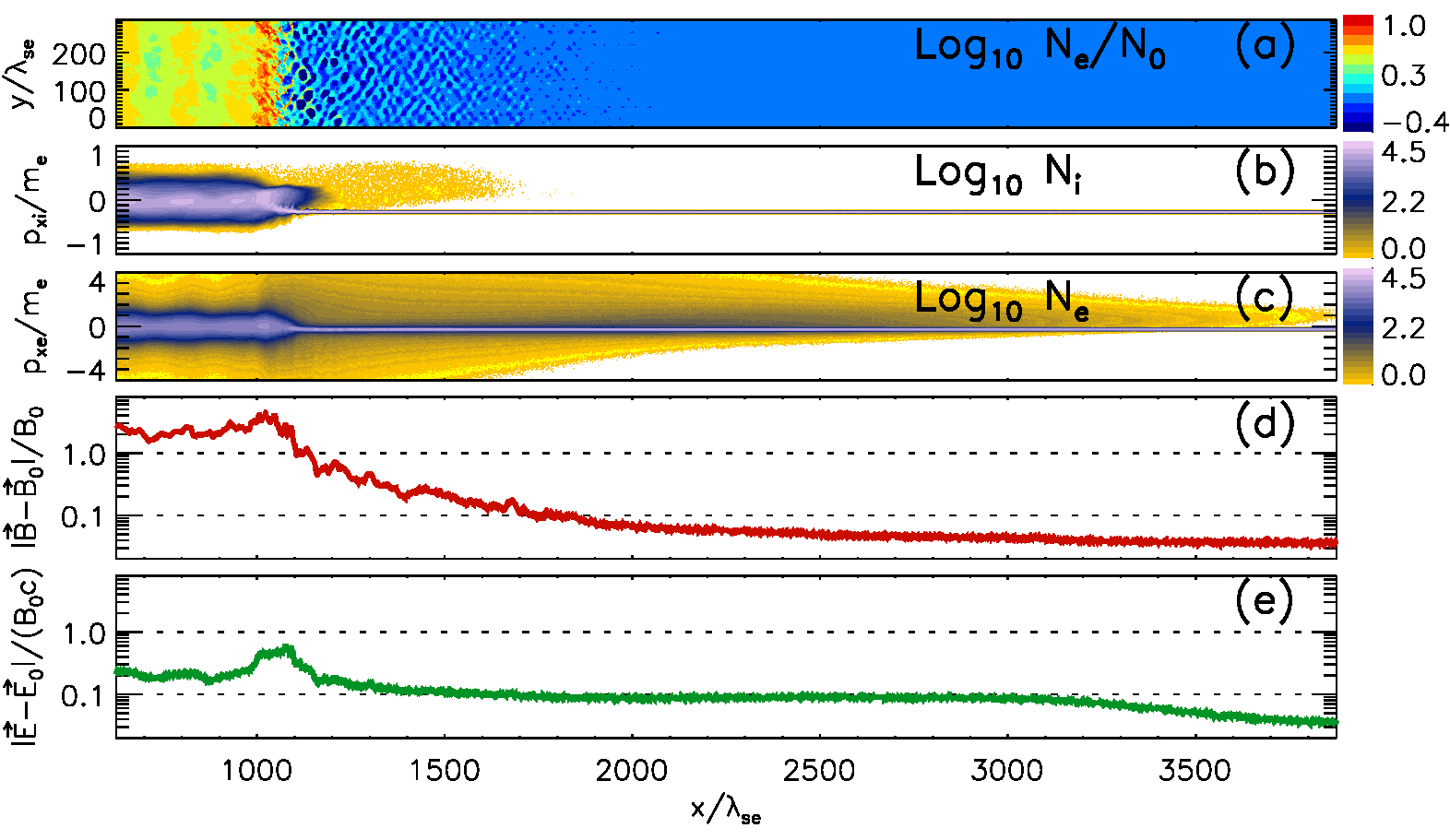}
    \caption{Map of the electron density (a), histograms of the $x$-$p_x$ phase-space density \abrev{($p_x$ is the particle momentum in x-direction)} for ions (b) and electrons (c), and profiles of the magnitude of the electric (d) and magnetic (e)  field at time $t = 18.1~\omci^{-1}$ \abrev{of the shock simulation}.}
    \label{figShockProfile}
\end{figure*}

The electron-density profile of the fully formed shock ($t = 18.1 \omci^{-1}$), shown in Fig.~\ref{figShockProfile}, suggests that the simulation domain can be divided into a downstream region ($x/\lse\lesssim1000$), the shock \mpo{transition} ($1000\lesssim x/\lse\lesssim1100$), \abrev{the foreshock that can be subdivided further according to the fluctuating field components ($1100\lesssim x/\lse\lesssim4000$), and the upstream plasma which contains the undisturbed plasma ($x/\lse\ \gtrsim 4000$).}
The downstream plasma exhibits turbulence on all length scales below $\lambda\approx 400\lse$ \mpo{that is} associated with \abrev{irregular shock evolution, also known as the shock self-reformation \citep{Wieland2016,Bohdan2017}, and electromagnetic waves transmitted into the downstream from the upstream region.} \mwein{The inner foreshock region ($1100\lesssim x/\lse\lesssim2000$)} is dominated by oblique waves with a wavelength \mpo{between $30\lse$ and $50\lse$ that have} a significant magnetic component ($|\vec B-\vec B_0|/B_0 \approx 0.1-1$, Fig.~\ref{figShockProfile}(d)). \abrev{In the outer foreshock shock ($2000\lesssim x/\lse\lesssim4000$)} the magnetic field is quiescent whereas strong electrostatic fluctuations \mpo{with a wavelength of $\lambda\approx (3-4)\lse$ and amplitude $|\vec E-\vec E_0|/(B_0c) \approx 0.1$ propagate along the $x$ axis (cf. Fig.~\ref{figShockProfile}(e))}.

\begin{figure*}[!t]
    \includegraphics[width=0.33\linewidth]{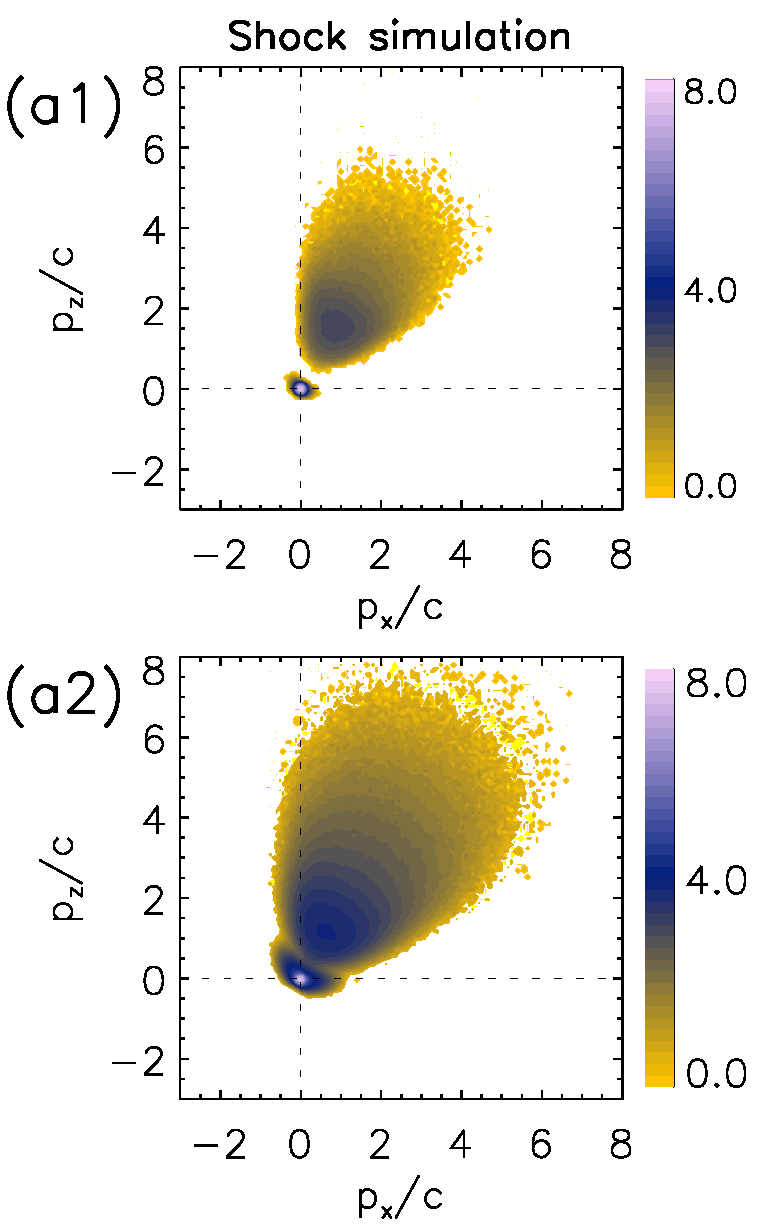}
    \includegraphics[width=0.33\linewidth]{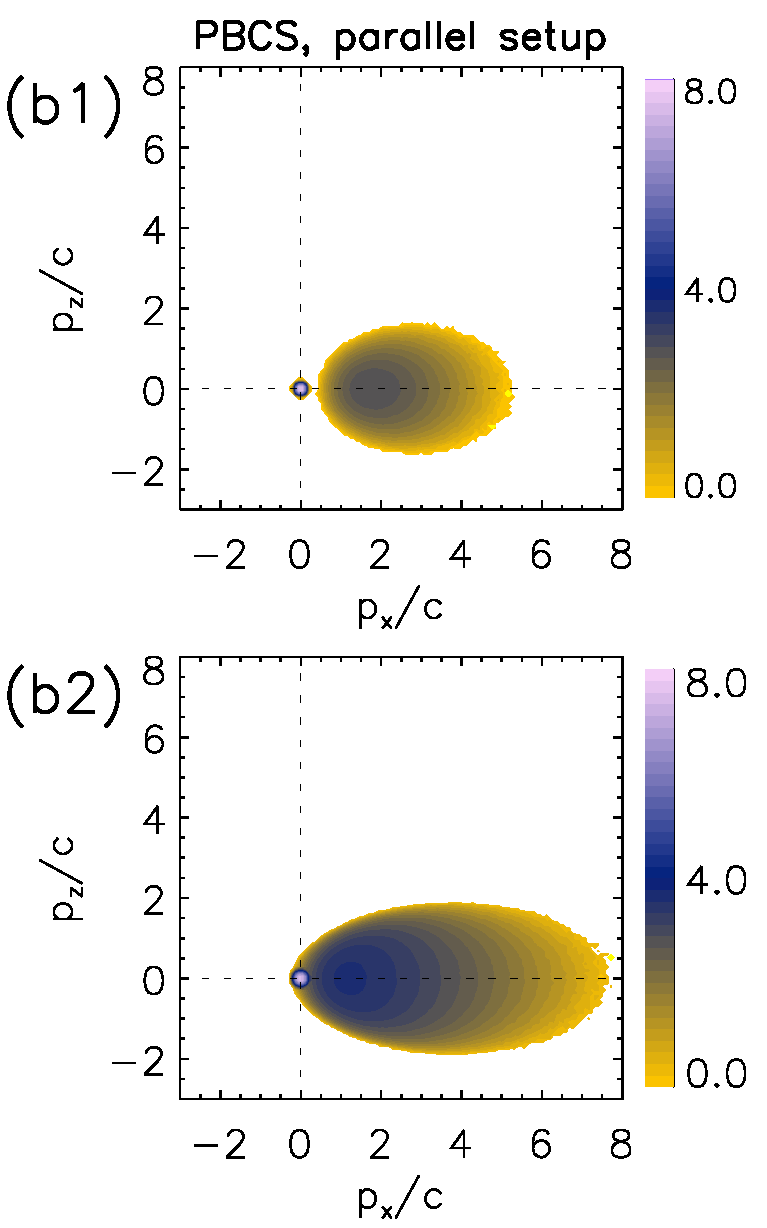}
    \includegraphics[width=0.33\linewidth]{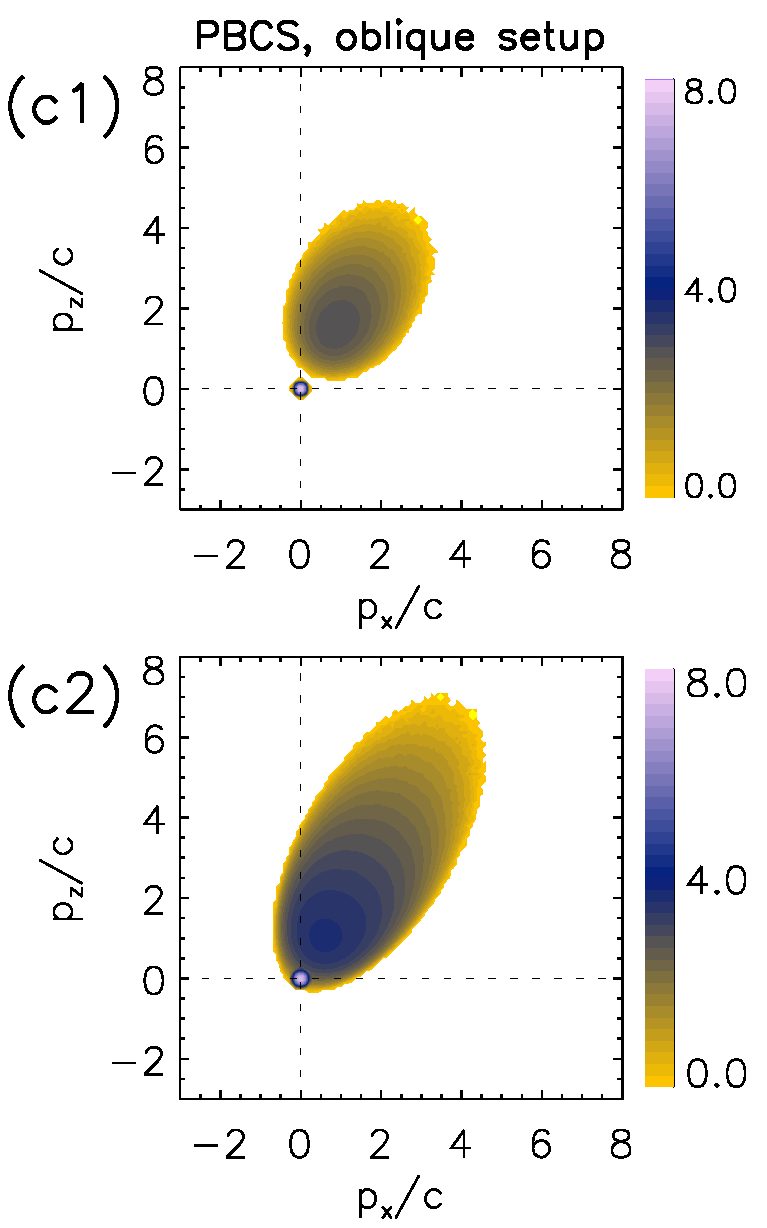}
    \caption{Electron momentum distributions projected on the $\vec\vsh$-$\vec B_0$ plane in representative sections of the (a1) \abrev{outer-foreshock} ($3250<x/\lse<3373$) and the (a2) \abrev{inner-foreshock} ($2500<x/\lse<2625$) regions of the shock simulation at $t = 18.1~\omci^{-1}$. \mpo{For comparison we also show the idealized} electron momentum distributions used in PBC simulations with parallel (b*) and oblique (c*) magnetic field for the \abrev{outer-foreshock} ((b1) and (c1)) and \abrev{inner-foreshock} ((b2) and (c2)) regions.}
    \label{figShockPS}
\end{figure*}

The \abrev{foreshock} region contains shock-reflected electrons \mpo{that stream through the background electrons with a mean velocity of $\vdr\approx0.9c$ in the entire upstream region, but the density of the reflected electrons significantly decreases farther upstream (Fig.~\ref{figShockProfile}(c)).} Their momentum spread also becomes smaller \mpo{at larger distance from the shock, suggesting} that electrons are more effectively scattered by the electromagnetic waves in the \abrev{inner foreshock} than by the exclusively electrostatic fluctuations farther from the shock.

Electrostatic and electromagnetic waves are excited only in the region where reflected electrons are present (see Sec.~\ref{sec:foreshocks-length}). Fig.~\ref{figShockPS}(a1) and (a2) shows the $p_x-p_z$ momentum distribution of electrons in two \abrev{regions which are prerequisite for generation of electrostatic and electromagnetic waves.} 
The selected regions lie where the excitation of oblique electromagnetic ($2500<x/\lse<2625$) and longitudinal electrostatic ($3250<x/\lse<3373$) waves begins to be measurable, but the waves are not yet strong enough to be visible in the density profile in Fig.~\ref{figShockProfile}(a).
We derive parameters describing the phase-space distributions of the upstream and the reflected electrons for the selected regions. In the next section we discuss results of simulations that are based on using derived plasma properties to \mpo{explore the} nature of observed waves and confirm their association with reflected electrons.

\section{Periodic-boundary-condition simulations (PBCS)} \label{sec:all-instab}

\begin{figure*}
    \centering
    \includegraphics[width=0.33\linewidth]{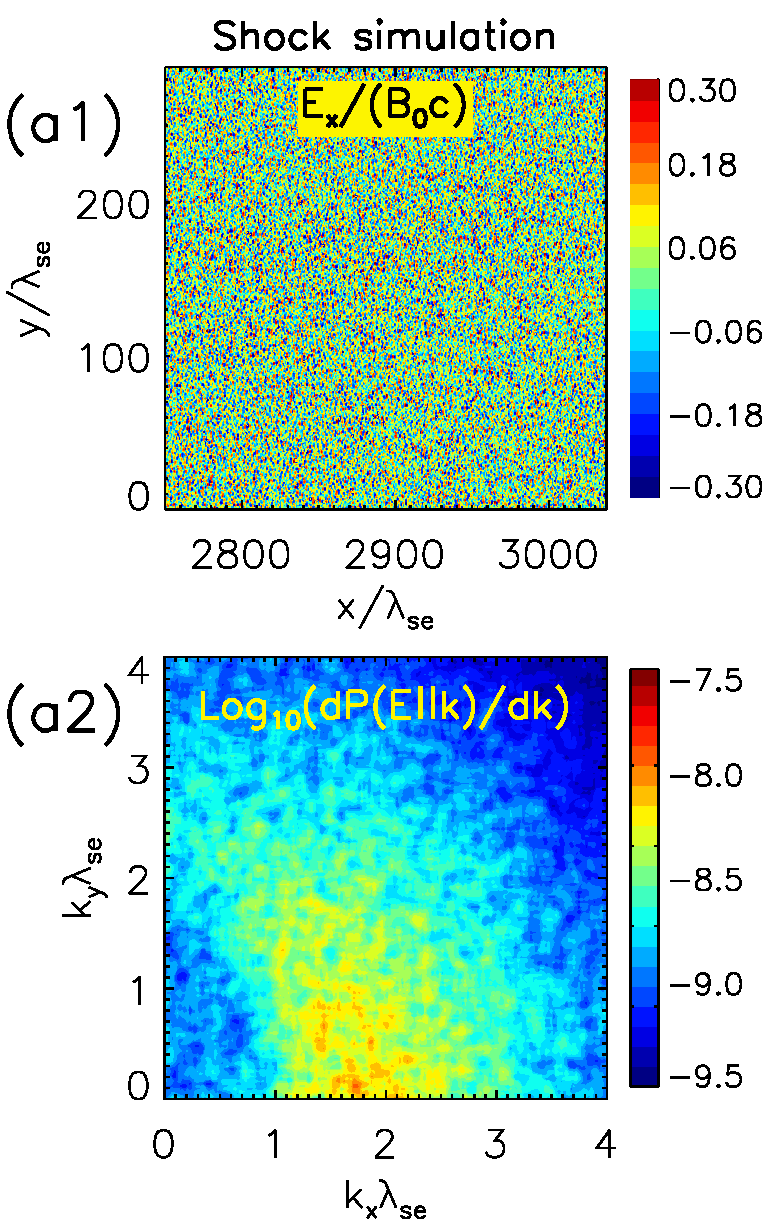}
    \includegraphics[width=0.33\linewidth]{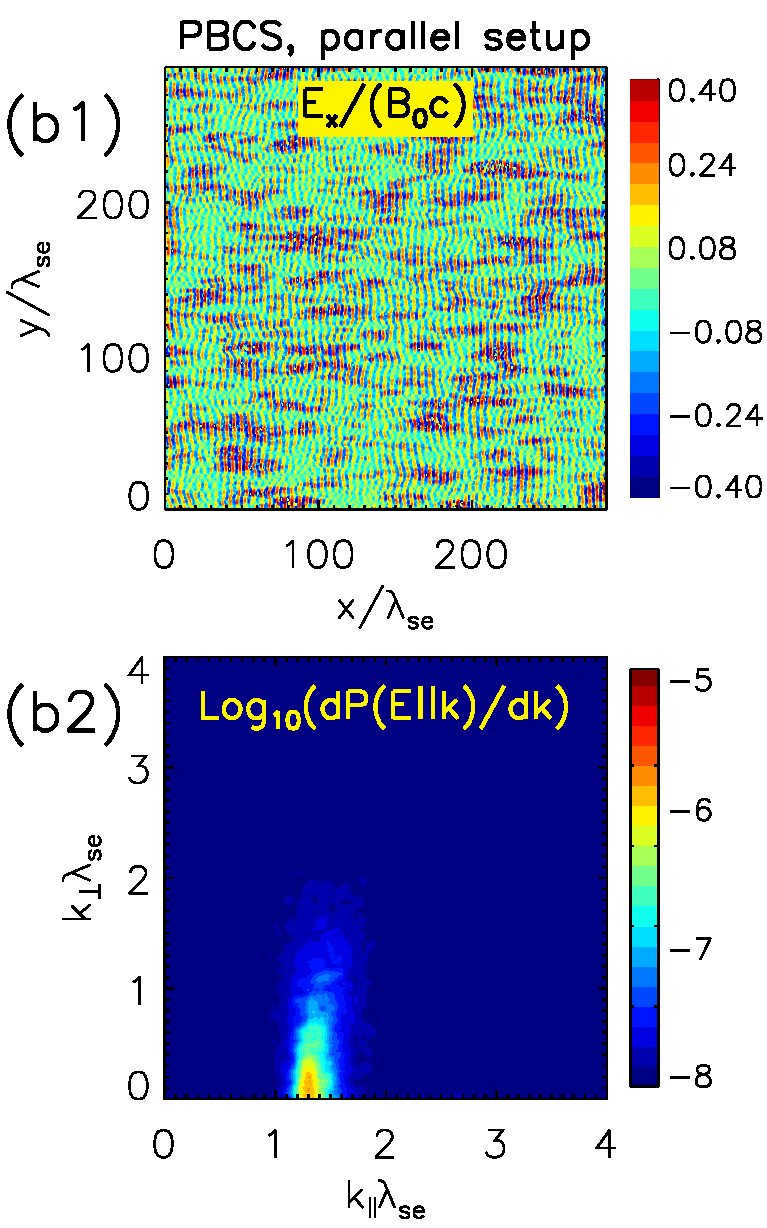}
    \includegraphics[width=0.33\linewidth]{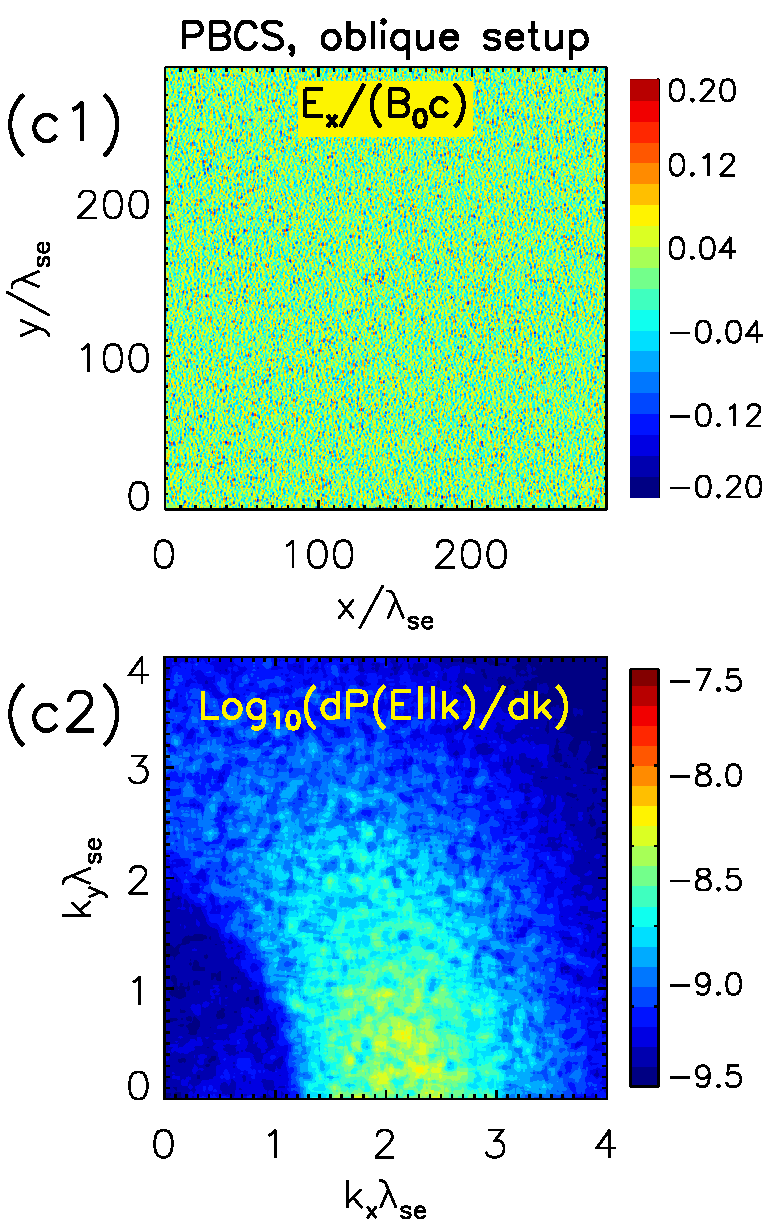}
    \caption{Electron-acoustic waves in the shock simulation (a*) and in the PBC simulations with parallel (b*) and oblique (c*) configurations. \mpo{The top panels display 2D maps} of $E_x$, and \mpo{the bottom panels give the} Fourier power spectrum of the electric field parallel to the in-plane component of the wave vector, $k \cdot E$. Snapshots are taken at $t = 18.1~\omci^{-1}$ for the shock simulation and at $t = 625~\ompe^{-1}$ for both PBC simulations.
    }
    \label{figure4_d_vert}
\end{figure*}

\subsection{PBC simulation setup}

Having identified two modes that are characteristic of the \abrev{outer and inner foreshock regions} in our shock simulation, respectively, we explore whether they can be locally excited by beam instabilities in the foreshock or are more likely to have propagated there from the shock front. Hence we run four PBC simulations and initialize the species with distributions similar to those we found in the two regions of interest in the shock simulation.

\mpon{For the initial momentum distribution of each particle species,} we use a bi-Gaussian distribution with the same second moment as in the representative regions. We then Lorentz-boost the distribution of the reflected electrons by their relative drift speed (see Fig.~\ref{figShockPS}). A return current between the background ions and electrons is not initialized, because there is no relative drift between these species in the upstream before they start interacting with the first reflected electrons. \mpon{Our} goal is to study the electrostatic waves that are generated as a result of this interaction. These waves will eventually trap electrons and establish a return current self-consistently.

To cover the widest range of propagation angles with respect to $\vec B_0$, we analyze a pair of PBC simulations for each \abrev{outer and inner foreshock}: one PBC simulation with a parallel magnetic field $\vec B_0 = B_0 \hat x$ (Fig.~\ref{figSetup}(b)) and one PBC simulation with an oblique magnetic field, $\vec B_0 = B_0 (0.5, 0, \sqrt{3}/2)$ (Fig.~\ref{figSetup}(c)), as in the shock simulation. \abrev{The drift velocity of reflected electrons is along the large scale magnetic field $\vec B_0$ and the absolute value is the same for parallel and oblique PBC simulations.} Each simulation extends over $376~\lse \times 361~\lse$, easily containing the wavelengths observed above, and is set in the upstream reference frame. The number of particles per cell is increased fivefold, $n_0=200$, for better statistical representation of the dilute beam of reflected electrons. \ab{The other plasma parameters \mpon{have the same value} as in the shock simulation.}
\abrev{Note that the two types of PBC simulations serve different purposes. PBCS with a parallel magnetic field are used to identify excited modes and compare simulations results with the linear dispersion analysis. PBCS with an oblique configuration demonstrate that PBCS results are consistent with the picture captured with the shock simulation.}

\subsection{\abrev{Outer-foreshock} conditions}

\begin{table}[]
    \centering
    \begin{tabular}{|c|c||c|c|}
    $n_b/n_0$ & $2.4\cdot10^{-3}$ 
    &  $A_b$ & 0.49 \\
    $u_b/c$   & 2.06 
    & $v_b/c$   & 0.90\\
    $u_{\mathrm{th}\parallel,0}/c$ & $3.14\cdot10^{-2}$
    & $v_{\mathrm{th}\parallel,0}/c$ & $3.14\cdot10^{-2}$\\
    $u_{\mathrm{th}\parallel,b}/c$ & 0.70 
    & $v_{\mathrm{th}\parallel,b}/c$ & 0.50 \\
    $\ompb'/\ompe$ & $3.20\cdot10^{-2}$& $\omce'/\omce$ & 0.43
    \end{tabular}
    \caption{Relevant parameters for the electron species in the \abrev{outer-foreshock} PBCS (with a normalized momentum $u=p/m$) and in the linear calculations with a bi-Maxwellian velocity distribution. Here $A_b=(v_{\mathrm{th}\perp,b}/v_{\mathrm{th}\parallel,b})^2$ denotes the beam temperature anisotropy and $\ompb'=(n_b/\gamma_b n_e)^{1/2}\ompe$ and $\omce'=\omce/\gamma$ denote the relativistically corrected beam plasma and gyrofrequency. See text for further definitions and other parameters.}
    \label{tabFarUpstream}
\end{table}

For the \abrev{outer-foreshock} region, we model the reflected electron beam observed in Fig.~\ref{figShockPS}(a1) as being homogeneously distributed throughout the simulation domain, with a beam density $n_b= 2.4\cdot10^{-3}~n_0$, drift velocity $|v_b|/c = 0.90$ along $\vec B_0$, and an anisotropic Maxwellian thermal spread in momentum space. The thermal velocities for the beam of reflected electrons in this model are $v_{\mathrm{th} \parallel,b} = 0.50~c$ and $v_{\mathrm{th} \perp,b} = 0.35~c$. In the simulation frame, the mean velocity of the background electrons with number density $n_0-n_b$ is zero, while their thermal spread is isotropic and equal to $v_{\mathrm{th},0} = 0.0314~c$ for each Cartesian direction. The background ions with number density $n_0$ also have zero mean drift and are initially in temperature equilibrium with the background electrons. The resulting electron momentum distributions are shown in Fig.~\ref{figShockPS}(b1) and~\ref{figShockPS}(c1) for parallel and oblique PBC simulations.

\begin{figure}
    \centering
    \includegraphics[width=0.99\linewidth]{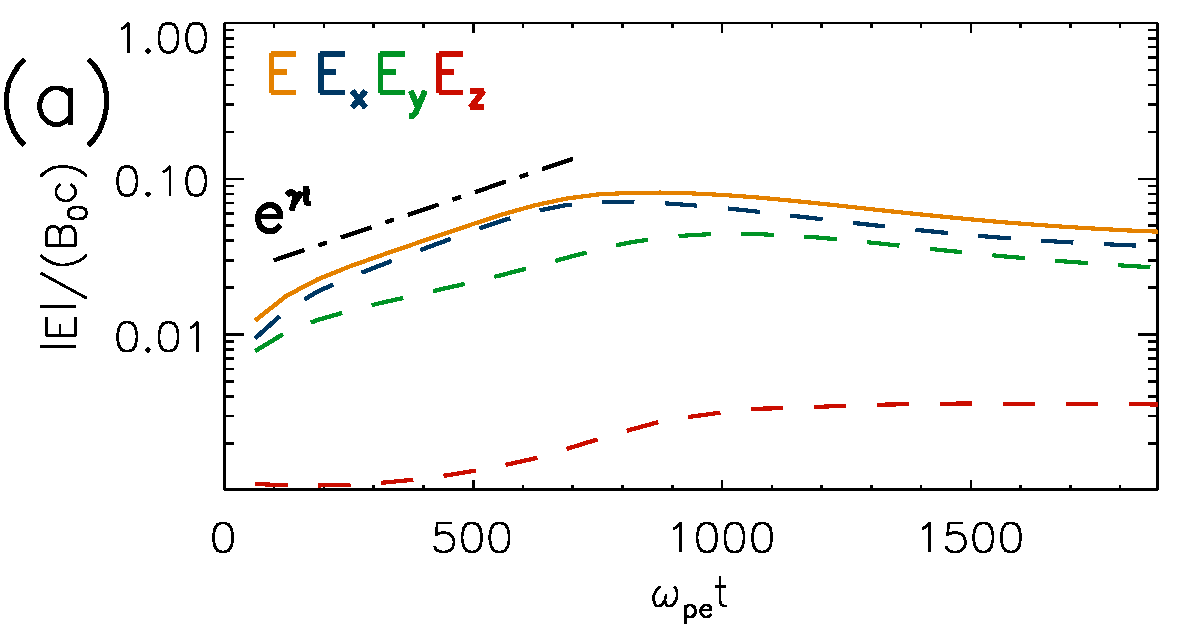}
    \includegraphics[width=0.99\linewidth]{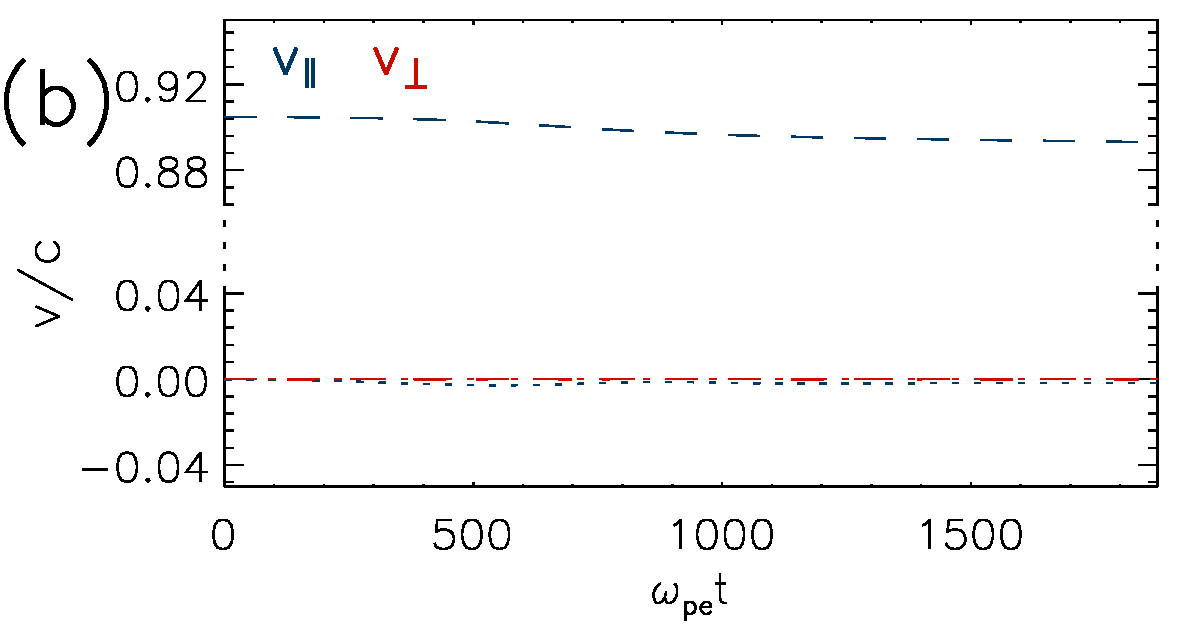}
    \includegraphics[width=0.99\linewidth]{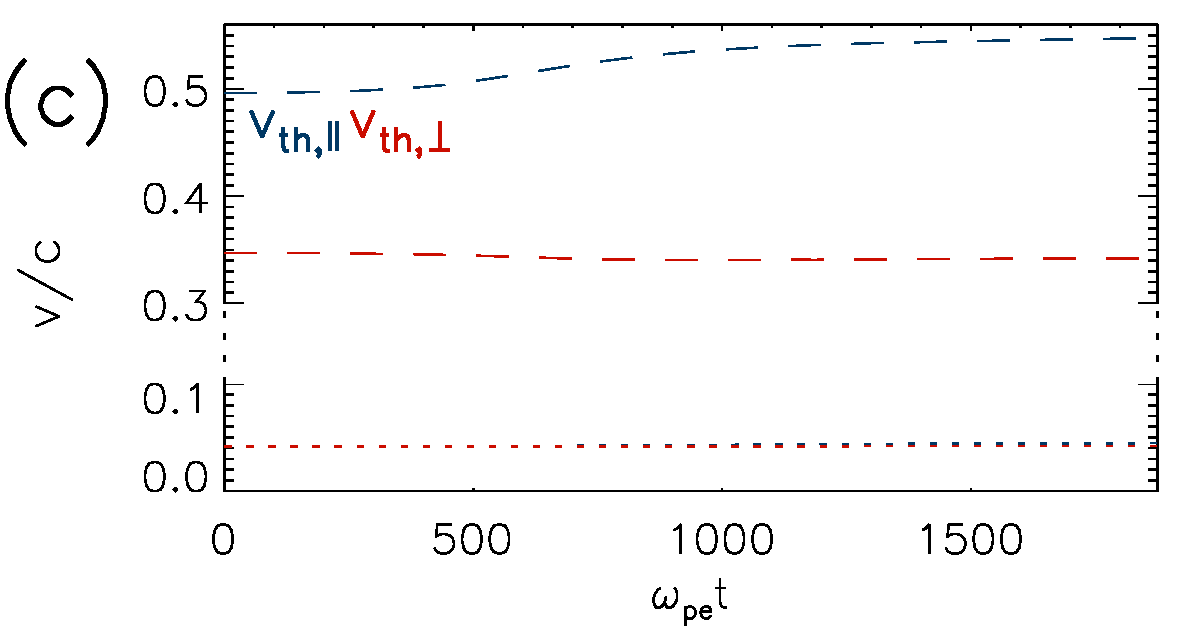}
    \includegraphics[width=0.99\linewidth]{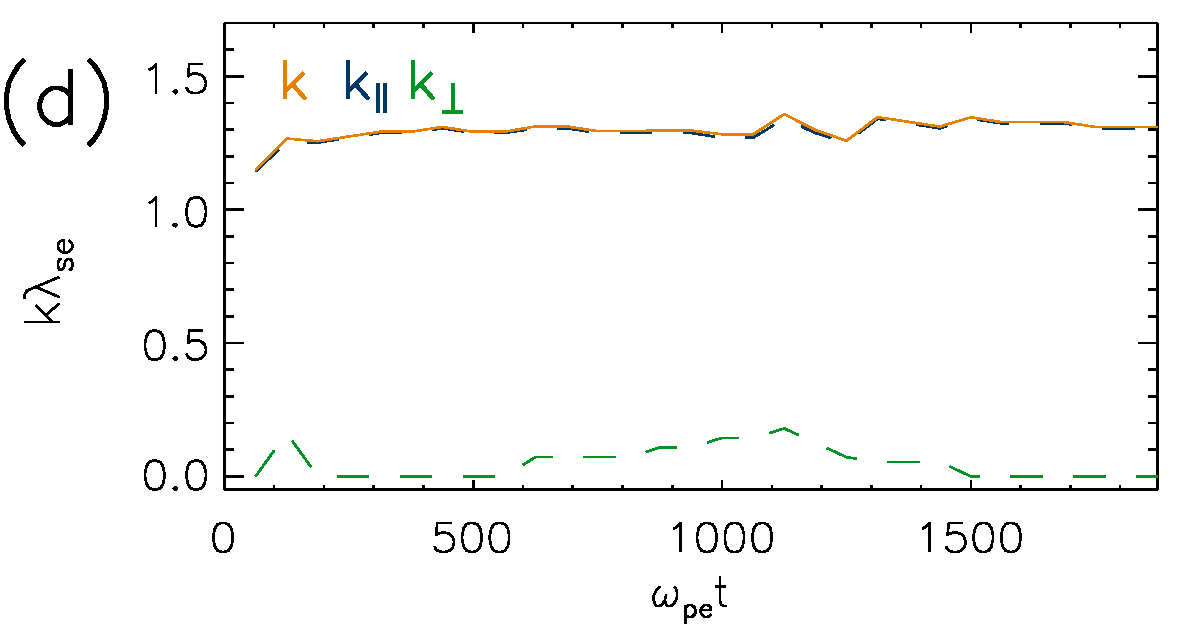}
    \caption{Evolution of the electric field (a), the bulk speed (b), the random velocity components of the electron (c), \abrev{and the most unstable mode (d)}, all averaged over the entire domain for the \abrev{outer-foreshock} PBCS in the parallel case. The dash-dotted line in the top panel indicates the peak growth rate $\gamma=2.5\cdot10^{-3}~\ompe$ obtained from the linear dispersion analysis. The dashed lines in panels (b) and (c) correspond to reflected electrons; dotted lines indicate background electrons.}
    \label{figElectrostaticPBC}
\end{figure}

Fig.~\ref{figure4_d_vert} compares two-dimensional maps of $E_x$ and power spectra of the wave-aligned electrostatic field, $E_{\parallel k}$, for the \abrev{outer-foreshock} region of the shock simulation (panels (a*)), the PBC simulation with the parallel magnetic field (panels (b*)), and the PBC simulation with the oblique magnetic field (panels (c*)). In all three simulations, $E_x$ quickly develops fluctuations that predominantly propagate in the $x$ direction. This direction corresponds either to the direction of the magnetic field (PBCS parallel) or to its projection onto the simulation plane (shock simulation and PBCS oblique). In the parallel PBCS the highest wave intensity is seen at $k_x \lse = 1.3$, while the peak in the spectral power shifts to larger wave numbers if the simulated plane is tilted with respect to the magnetic field.

For the PBCS in the parallel configuration, the electric field continues growing exponentially until $t = 700~\ompe^{-1}$ (Fig.~\ref{figElectrostaticPBC}(a)). As the instability operates, the difference in bulk velocity between the reflected and the background electrons becomes slightly smaller (Fig.~\ref{figElectrostaticPBC}(b)). More clearly visible is the parallel heating of the reflected electrons from $v_{\mathrm{th}\parallel,b}=0.50~c$ to $0.55~c$ during the exponential-growth phase (Fig.~\ref{figElectrostaticPBC}(c)), which continues even after the electric field has saturated. During the exponential phase, the electric field grows at a rate of about $2.5\cdot10^{-3}~\ompe$, as indicated by the dash-dotted line in Fig.~\ref{figElectrostaticPBC}(a).

\begin{figure}
    \centering
    \includegraphics[width=.99\linewidth]{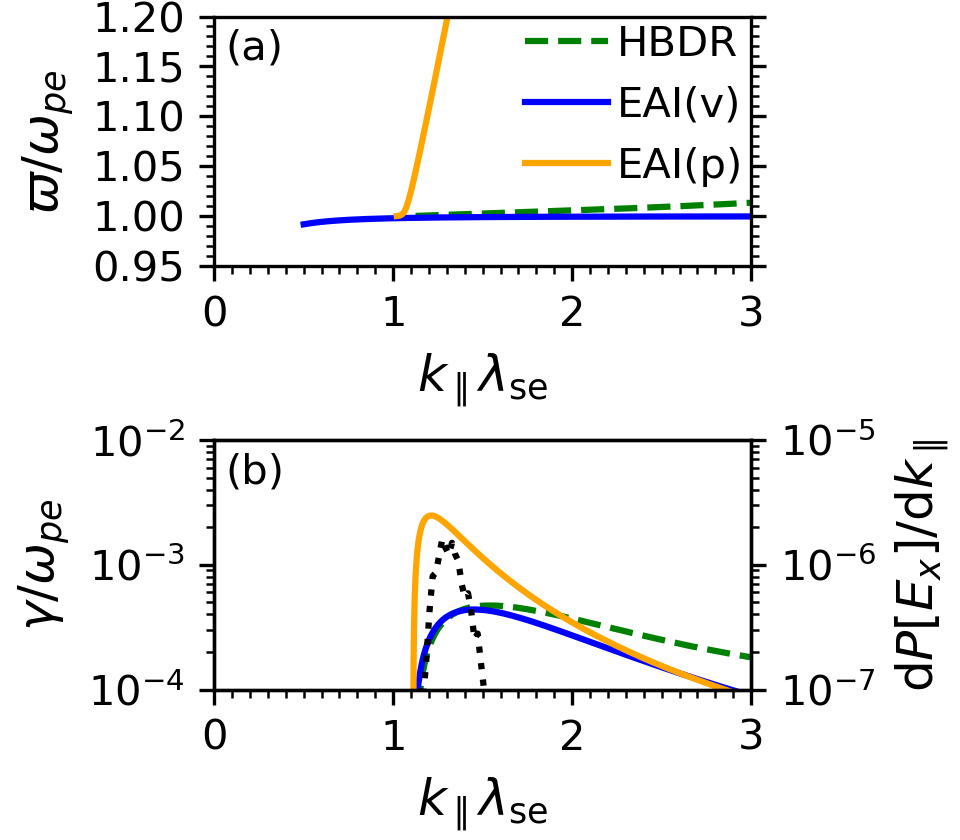}
    \caption{Angular frequency, $\varpi$ (a), and growth rate, $\gamma$ (b), of the electrostatic instability for \abrev{outer-foreshock} parameters and the parallel PBCS setup, calculated numerically for the bi-Maxwellian hot-beams dispersion relation (green dashed) and analytically for the electron-acoustic instability with Gaussian velocity (blue) and Gaussian momentum distributions (orange). For comparison, the dotted line (and right-hand axis) give the power spectrum of $E_x$ at its peak, $\ompe~t=700$.}
    \label{figElectrostaticGamma}
\end{figure}

The excitation of an electrostatic instability is also consistent with a linear dispersion analysis of the electron-beam configuration. We find numerical solutions of the hot-beams dispersion relation with {\scshape Whamp} \citep{Ronnmark82}, which employs various approximations of the Fried--Conte plasma dispersion function. Because this function is obtained from the dispersion relation for bi-Maxwellian beams, we model the three particle species as anisotropic Gaussian distributions in velocity space:
\begin{equation}
f_s(v_\parallel,v_\perp) = \frac{\exp\left\{-v_{\mathrm{th} \parallel,s}^{-2} \left[ (v_\parallel-v_{s})^2 + v_\perp^2/A_s \right] \right\}}{\pi^{3/2}~v_{\mathrm{th} \parallel,s}^3~A_s}.
\label{eqnVeloGauss}
\end{equation}
Here $v_s$ denotes the bulk drift speed of the species $s$ along the background magnetic field, $A_s = (v_{\mathrm{th} \perp,s}/v_{\mathrm{th} \parallel,s})^2$ denotes the temperature anisotropy, and the species index $s\in\{0,b,i\}$ refers to the background electrons, the reflected-beam electrons, and the background ions. Although the relativistic electron distributions shown in Fig.~\ref{figShockPS} deviate significantly from a Gaussian in velocity, and although the initial distributions in the PBCS are bi-Gaussian in momentum as opposed to velocity, we can already achieve some agreement with standard linear theory by only keeping the first two moments for each species and increasing the mass of the reflected electrons by their bulk Lorentz factor $\gamma_b$. This correction for the relativistic mass of the reflected electrons reduces the classical definitions of the beam plasma and cyclotron frequency by $\gamma_b^{1/2}$ and $\gamma_b$, respectively, and will become important for determining the gyroresonant frequencies of oblique whistler waves. The corrected frequencies $\ompb'$ and $\omce'$, together with other parameters for the \abrev{outer-foreshock} distribution crucial for the following calculations, are summarized in Table~\ref{tabFarUpstream}.

Fig.~\ref{figElectrostaticGamma} shows the angular frequency and the growth rate of the electrostatic modes that are linearly unstable in this bi-Maxwellian calculation, using the standard hot-beams dispersion relation (dashed green). The fastest-growing mode  reaches a growth rate of $5\cdot10^{-4}~\ompe$, only about a fifth of the rate that we observe for the exponential phase of the electric-field growth in the simulation shown in Fig.~\ref{figElectrostaticPBC}(a). Yet the wave number of the peak predicted by the bi-Maxwellian calculation ($k_\parallel \lse = 1.5$) is in good agreement with the power spectrum of $E_x$ when it saturates, indicated by the dotted line in Fig.~\ref{figElectrostaticGamma}(b).

The blue lines in Fig.~\ref{figElectrostaticGamma} indicate the analytic solution for the electron-acoustic instability \citep{Gary87} driven by a bi-Maxwellian beam of reflected electrons with mass $\gamma_b m_e$, drift velocity $v_b$, and thermal velocity $v_{\mathrm{th} \parallel,b}$. This solution represents the standard electron-acoustic instability with a Gaussian distribution for the beam velocity. It is clearly an excellent match for the linear hot-beams dispersion relation, which means that it also underestimates the growth rate observed in the PBCS.

Similar to the electrostatic two-stream instability, which is excited by a cold beam of charged particles, the electron-acoustic instability is driven by Landau resonance between plasma waves in a cold background plasma and a beam of fast electrons with a large thermal spread. Because of this resonant character, the electron-acoustic instability is particularly sensitive to the distribution of $v_\parallel$. In order to find a more accurate prediction for our PBCS, we also find the analytic solution of a modified electron-acoustic instability, assuming a Gaussian momentum distribution for the beam and using the same parameters with which we initialize the PBCS. The growth rate then reads (see Appendix~\ref{appEAIp}):
\begin{equation}
    \frac{\gamma(k)}{\ompe}=-\sqrt{\frac{\pi}{2}}\ \frac{\ompb^2}{k^2}\ \frac{u_{\mathrm{res}}(k)-u_0}{u_{\mathrm{th}}^3} \exp\left[-\frac12 \left(\frac{u_{\mathrm{res}}(k)-u_0}{u_{\mathrm{th}}}\right)^2\right].
    \label{eqnRelElAcoustic}
\end{equation}
Here $u=p/m$ is the normalized relativistic momentum and the resonant momentum $u_{\mathrm{res}}$ follows from $u_{\mathrm{res}}/(1+u_{\mathrm{res}}^2/c^2)^{1/2}=\ompe/k$. As in the PBCS initialization, we use $\ompb^2=2.4\cdot10^{-3}~\ompe^2$, $u_0=2.06c$, and $u_{\mathrm{th}}=0.7c$. 

The modified calculation assumes a much higher phase-space density at velocities $v\lesssim c$ than the standard Maxwellian-beam calculation, which spreads some of the distribution into the unphysical velocity range $v>c$. This modification results in a larger angular frequency and growth rate for the electron-acoustic instability (orange lines in Fig.~\ref{figElectrostaticGamma}). The fastest-growing mode ($k_\parallel \lse \approx 1.2$) reaches a growth rate of $2.5\cdot10^{-3}~\ompe$, in much better agreement with the simulations. \mwein{As indicated by Fig.~\ref{figElectrostaticPBC}d and the dotted line in Fig.~\ref{figElectrostaticGamma}b, the spectral power of the electrostatic field peaks at almost the same wave number ($k_\parallel~\lse\approx1.3$) in the parallel PBCS.}

While this improvement confirms that we observe the (modified) electron-acoustic instability in the longitudinal direction, generalizing the Gaussian-momentum modification to arbitrary propagation directions would be extremely difficult. Thus we will continue to use the numerical solution of the hot-beams dispersion relation as reference for modes with a transverse component.

At oblique wave vectors, the fastest-growing electron-acoustic mode has a larger wave number and a smaller growth rate than it has parallel to the field, which qualitatively agrees with the power spectra of our shock simulation (Fig.~\ref{figure4_d_vert}(a2)) and the PBC simulation in the oblique configuration (Fig.~\ref{figure4_d_vert}(c2)). Specifically at an angle $\theta_{Bn}=60^\circ$, the numerically derived growth rate of the electrostatic instability peaks at $|k| \lse = 3.5$ with $\gamma = 9.7\cdot10^{-5}~\ompe$, \abrev{which is in rough agreement with PBC oblique simulations}. This prediction underestimates by almost a factor of two the wavelength of the fastest-growing mode: the electrostatic mode in the shock simulation and the PBCS with an oblique field peaks at $|k| \lse \approx 2$. We attribute this difference again to the distribution of reflected particles in the simulation, which is initially Gaussian in momentum space, but not in velocity space as the numerical model assumes.

The linear dispersion analysis of the \abrev{outer-foreshock} conditions additionally predicts an electromagnetic instability at oblique angles, similar to the whistler waves described below. However, the growth rate for this oblique mode is so small, below $2\times10^{-5}~\ompe$ throughout wave-vector space, that its amplitude would remain negligible until the shock front arrives. Hence, to explain the oblique waves observed in the shock simulation, we must consider the denser beam of reflected electrons in the \abrev{inner foreshock}.

\subsection{\abrev{Inner-foreshock} conditions}

\begin{table}[]
    \centering
    \begin{tabular}{|c|c||c|c|}
    $n_b/n_0$ & $2.3\cdot10^{-2}$ 
    &  $A_b$ & 0.37 \\
    $u_b/c$   & 1.95 
    & $v_b/c$   & 0.89\\
    $u_{\mathrm{th}\parallel,0}/c$ & $3.14\cdot10^{-2}$
    & $v_{\mathrm{th}\parallel,0}/c$ & $4.18\cdot10^{-2}$\\
    $u_{\mathrm{th}\parallel,b}/c$ & 1.03 
    & $v_{\mathrm{th}\parallel,b}/c$ & 0.61 \\
    $\ompb'/\ompe$ & $1.02$& $\omce'/\omce$ & 0.45
    \end{tabular}
    \caption{Relevant parameters for the electron species in the \abrev{inner-foreshock} PBCS and in the linear calculations with a bi-Maxwellian velocity distribution. See Table~\ref{tabFarUpstream} and text for definitions.}
    \label{tabNearUpstream}
\end{table}

Thus we set up a second couple of PBC simulations, this time with a phase-space distribution similar to that in Fig.~\ref{figShockPS}(b) (see Table~\ref{tabNearUpstream}). The number density of the reflected-electron beam increases almost tenfold to $n_b=2.3\cdot10^{-2}~n_0$. This close to the shock, the thermal spread of the reflected electrons is slightly larger, thus $v_{\mathrm{th} \parallel,b} = 0.61~c$ and $v_{\mathrm{th} \perp,b} = 0.36~c$, whereas their mean drift is slightly smaller than before, $|v_b|/c = 0.89$. The background electrons have been scattered by sufficient electrostatic waves in the \abrev{outer foreshock} that their thermal spread has increased to $v_{\mathrm{th},0} = 0.0418~c$. The ion population is initially isothermal with the background electrons.

\mwein{We do not include the very small population of shock-reflected ions (about $10^{-4}~n_0$) that appears in the inner-foreshock region of the shock simulation. According to our linear estimates, the growth rate of the Buneman instability driven by these reflected ions is over two orders of magnitude smaller than the growth rate for the electromagnetic instability that we observe. While shock-reflected ions are crucial for the waves generated inside the shock foot, we can ignore them for our models of the inner foreshock.}

\begin{figure*}
    \centering
    \includegraphics[width=0.33\linewidth]{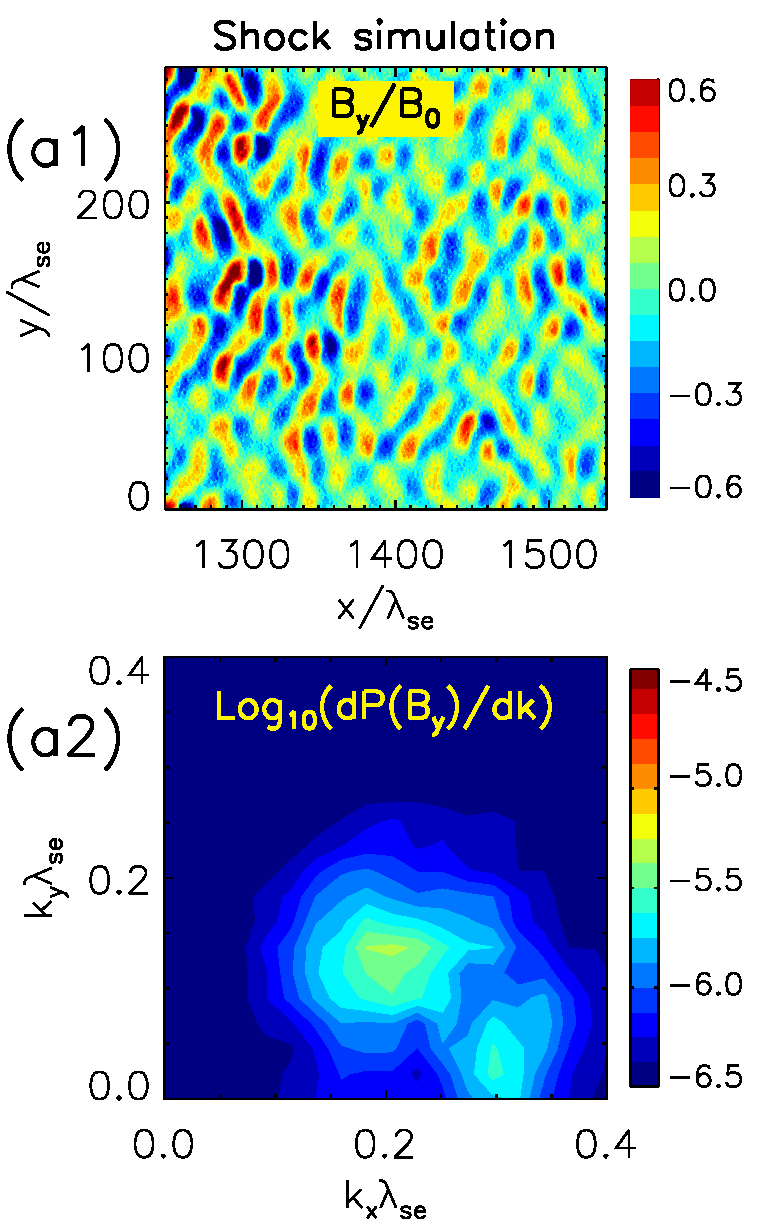}
    \includegraphics[width=0.33\linewidth]{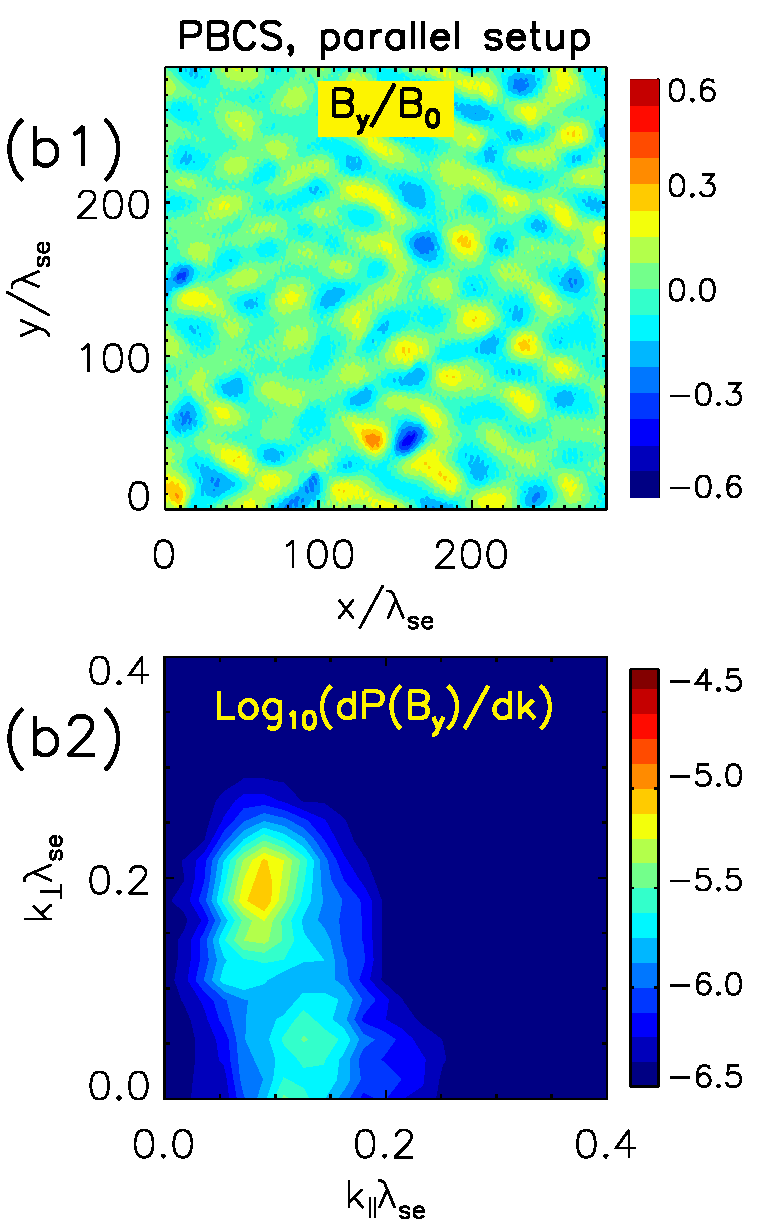}
    \includegraphics[width=0.33\linewidth]{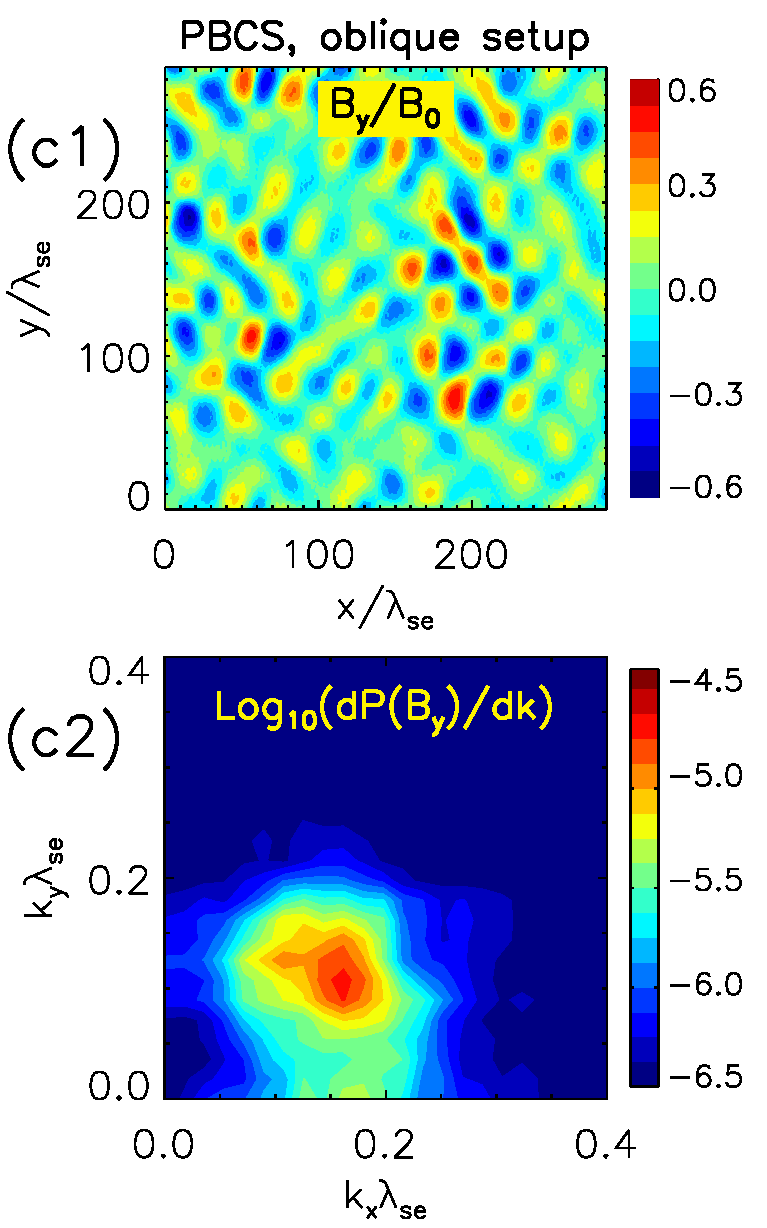}
    \caption{Whistler waves in the shock simulation (a*) and in the PBC simulations with parallel (b*) and oblique (c*) configurations. The top row gives 2D maps of $B_y$, and below that one finds Fourier power spectra of the magnetic field parallel to the in-plane component of the wave vector, $k \cdot B$ (bottom). Snapshots are taken at $t = 18.1~\omci^{-1}$ for the shock simulation and at $t = 8750~\ompe^{-1}$ for both PBC simulations.}
    \label{figWhistlerProfile}
\end{figure*}

In Fig.~\ref{figWhistlerProfile}, we compare maps of the perpendicular magnetic field, $B_y$, and its Fourier spectrum for the \abrev{inner-foreshock} region of the shock simulation (panels (a*)) with those for the two PBC simulations (panels (b*) and (c*)). Oblique fluctuations with a wavelength of about $30~\lse$ dominate the magnetic field in each case, although the fastest-growing mode is located at a larger $k_y (\approx0.19~\lse^{-1})$ and a smaller $k_x (\approx0.09~\lse^{-1})$ in the parallel PBCS \mwein{than in the oblique PBCS, where the peak lies at $(k_x\lse,k_y\lse)\approx(0.16,0.11)$. However, this is likely a projection effect due to the obliquely oriented simulation plane in the latter simulation. Transforming the wave vector where the spectrum of the oblique PBCS peaks to coordinates relative to the out-of-plane magnetic field, we find that the spectral maximum lies at $k_\parallel\lse\approx0.18$ along $\vec B_0$ and $k_\perp\lse\simeq0.08$ perpendicular to it, and thus at almost the same obliquity angle and wavenumber as for the parallel PBCS.} Unlike the electron-acoustic mode above, which grew significantly faster in the setup with an in-plane field, the spectral power for these oblique magnetic waves peaks at comparable values in both PBC simulations.

\begin{figure}
    \centering
    \includegraphics[width=0.99\linewidth]{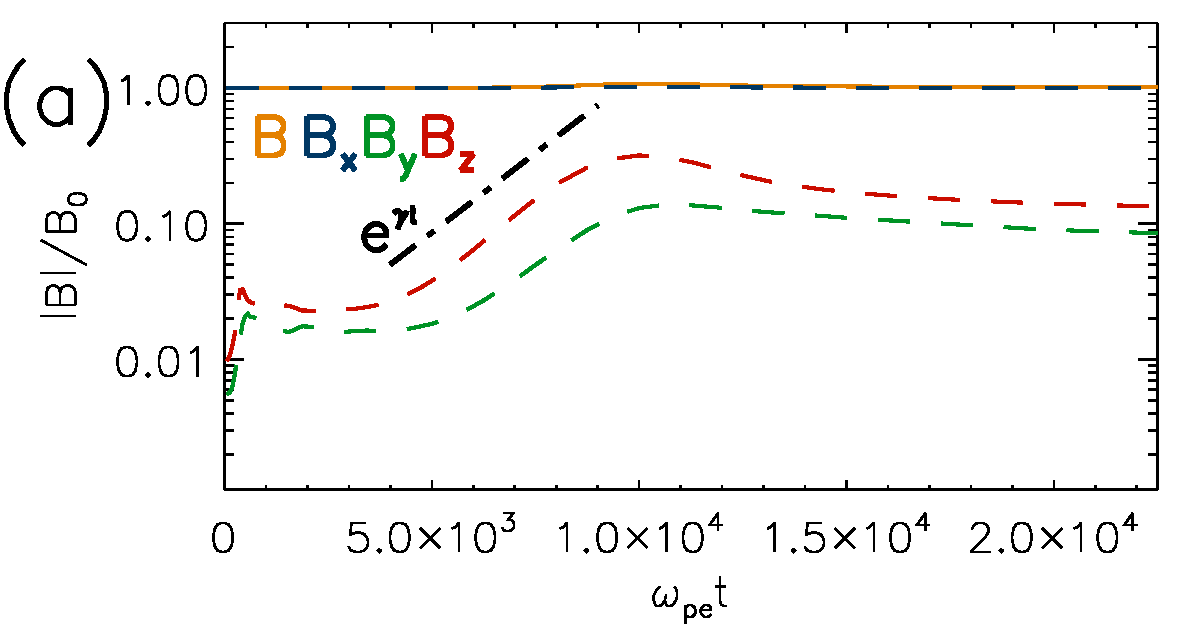}
    \includegraphics[width=0.99\linewidth]{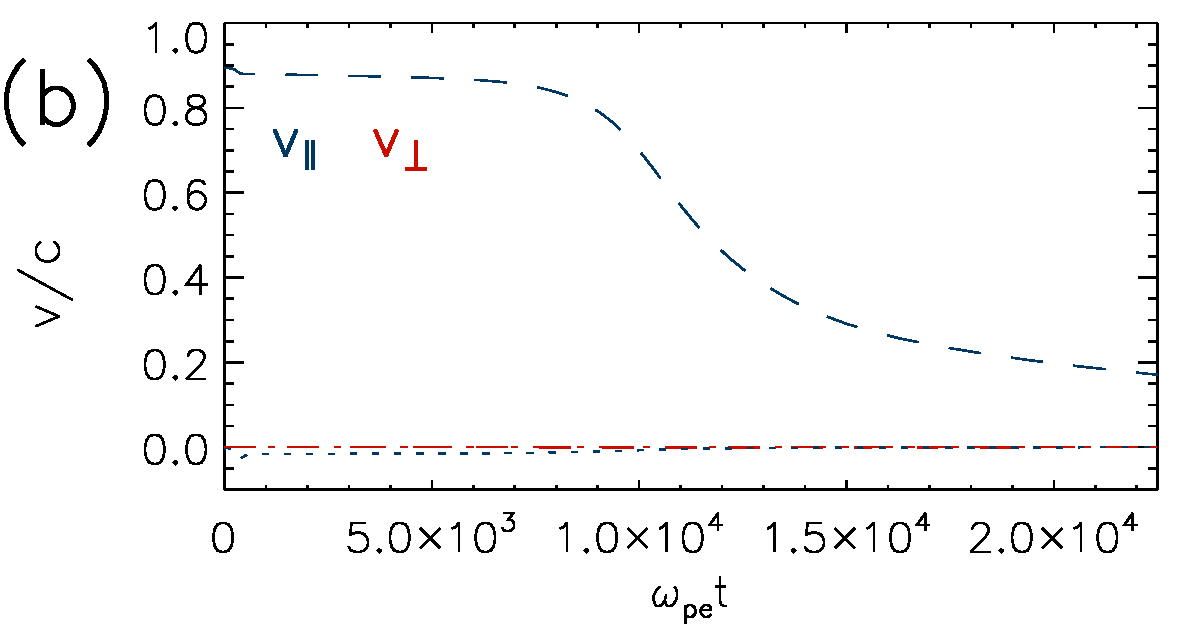}
    \includegraphics[width=0.99\linewidth]{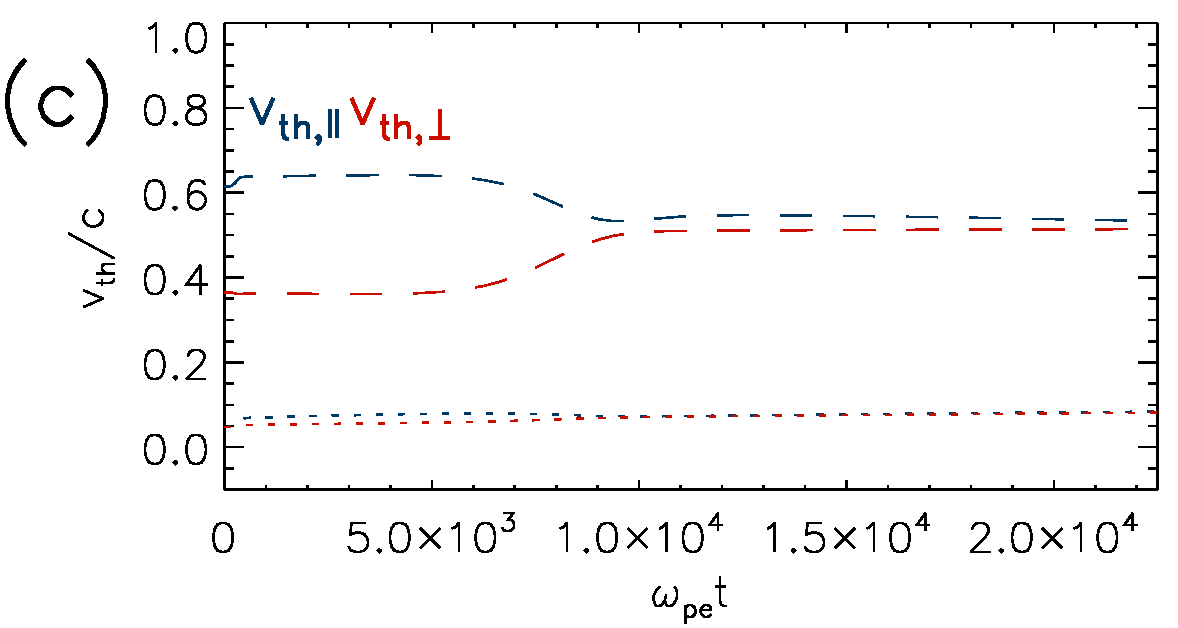}
    \includegraphics[width=0.99\linewidth]{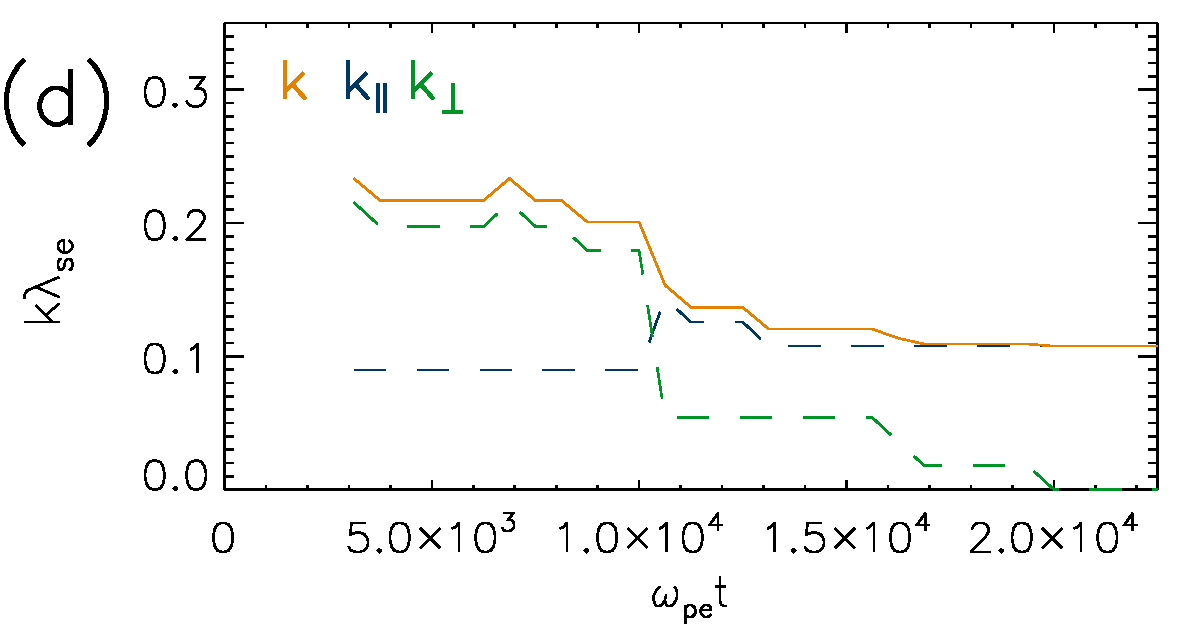}
    \caption{Evolution of magnetic field (a), bulk speed (b), the random velocity components of the electron (c), \abrev{and the most unstable mode (d)}, all averaged over the entire domain for the \abrev{inner-foreshock} PBCS with the parallel configuration. The dash-dotted line in the top panel indicates the peak growth rate $\gamma=0.009|\Omega_e|$ obtained in a linear dispersion analysis (see text). In panels (b) and (c) dashed lines refer to reflected electrons and dotted lines correspond to upstream electrons.}
    \label{figWhistlerPBC}
\end{figure}

The evolution of the magnetic-field components as well as the bulk drift and thermal velocity of both electron species in the PBCS-parallel are shown in Fig.~\ref{figWhistlerPBC}. Not depicted is the rapid but brief growth of electrostatic waves at the beginning of the simulation, similar to the previously discussed \abrev{outer-foreshock} simulation. Shortly after the electron-acoustic instability has saturated, the perpendicular magnetic field in \abrev{inner-foreshock} conditions grows exponentially and saturates within about $10^4~\ompe^{-1}$. During this stage of the simulation, the \ab{thermal \mpon{(random kinetic)} energy of} the reflected electrons increases by 100\%, after barely changing during the short period of electrostatic fluctuations. At the same time, the bulk speed of reflected electrons begins to decline from the original $v_b/c=0.89$ to $v_b/c\approx0.2$ at the end of the PBC simulation.

The fast scattering of the reflected electrons suggests that they are resonant with the oblique electromagnetic waves. To further investigate this putative resonance, we again find numerical solutions of the linear dispersion relation for hot beams, modeling the reflecting electrons with equation~(\ref{eqnVeloGauss}) using $v_{\mathrm{th}\parallel,b}=0.61c$, $A_b=0.37$, and $\gamma_b=2.2$. The fastest-growing electromagnetic mode in this calculation has a growth rate of $\gamma\simeq 9.0\cdot10^{-3}~|\omce|\simeq 5.4\cdot10^{-4}~\ompe$, in good agreement with the growth of the magnetic field \abrev{both in PBCS-parallel (Fig.~\ref{figWhistlerPBC}(a)) and PBCS-oblique}. While the perpendicular wave number of this mode, $k_\perp~\lse=0.17$, and the spectral peak seen in our PBCS with in-plane field match well, the linearly predicted parallel wave number is too small ($k_\parallel\lse=0.05$).

\begin{figure}
    \centering
    \includegraphics[width=.99\linewidth]{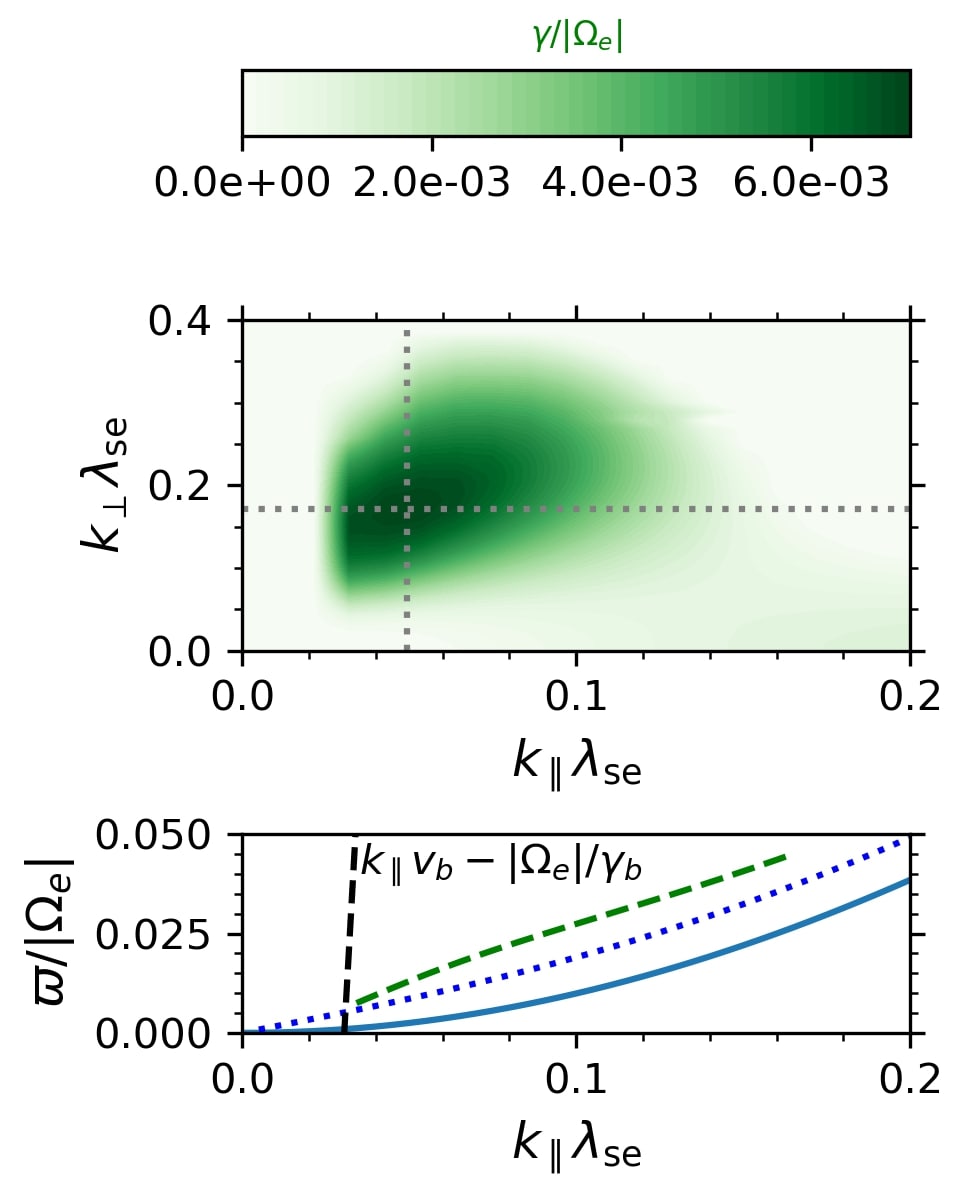}
    \caption{Top: growth rate of the oblique whistler mode, calculated with the bi-Maxwellian hot-beams dispersion relation for the \abrev{inner-foreshock} conditions. Bottom: angular frequency of the growing whistler waves from the hot-beams calculation (green dashed) compared with the analytic cold-whistler dispersion relation at $k_\perp=0$ (blue solid) and $k_\perp=0.017~\lambda_{se}^{-1}$ (blue dotted). Wave excitation is strongly suppressed where the Doppler-shifted gyrofrequency of the reflected electrons (black dashed) is smaller than the whistler frequency.}
    \label{figWhistlerGamma}
\end{figure}

Like whistler waves, the mode that is excited by the beam in the numerical model is right-hand polarised, and its angular frequency is close to that of oblique whistler waves in the background plasma. Fig.~\ref{figWhistlerGamma} compares the frequency of the numerical solution for $k_\perp\lse=0.17$ (green dashed line) with the solutions $\varpi_W$ of the cold-whistler dispersion relation for $k_\perp\lse=0.17$ (blue dotted) and for $k_\perp=0$ (blue solid),
\begin{equation}
\left(\frac{ck}{\varpi_W}\right)^2 = \frac{\ompe^2}{\varpi_W~\left( |\omce| \cos\theta - \varpi_W\right)},
\end{equation}
where $\cos\theta=k_\parallel/|k|$. Note that the green line only begins at $k_\parallel~\lse\approx0.03$: this oblique-whistler mode is stable for smaller parallel wave numbers because the beam can only excite fluctuations with sufficiently small phase speed, i.e.\ with frequencies smaller than the condition for first-order anomalous gyroresonance, $\varpi_W < \omres^{(1)}$. We define the $\ell$-th order anomalous gyroresonance for an electron beam with parallel velocity $v_b$ and Lorentz factor $\gamma_b$ as
\begin{equation}
\omres^{(\ell)} = k_\parallel v_{b} - \ell~\frac{|\omce|}{\gamma_b}.
\label{eqnOmegaRes}
\end{equation}
For the oblique-whistler instability in our simulations, anomalous gyroresonances up to third order play a role. The first-order gyroresonance marks the onset of the instability, the second-order gyroresonance at $k_\parallel\lse\approx0.05$ coincides with the fastest-growing mode in the linear model, and the third-order gyroresonance at $k_\parallel\lse\approx0.09$ lies remarkably close to the spectral peak of the magnetic field in the PBC simulations. This discrepancy between linear theory and simulation is again likely due to the difference between the Gaussian velocity or momentum distribution of the reflected electrons, which however only has a negligible effect on the perpendicular wave number of the fastest-growing mode. Of course, nonlinear effects like the evolution of the electron distribution may also account for some of the discrepancy from the linear prediction.

\begin{figure}
    \centering
    \includegraphics[width=.99\linewidth]{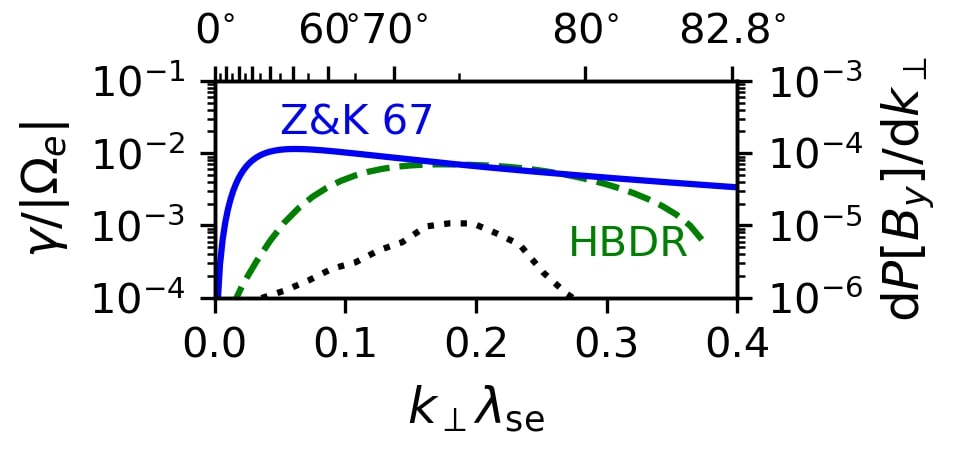}
    \caption{Left axis: growth rate of the oblique whistler mode at its theoretical peak, $k_\parallel \lambda_{se}=0.05$, calculated with the bi-Maxwellian hot-beams dispersion relation for the \abrev{inner-foreshock} conditions (green dashed) and with the cold-beams approximation of Zayed \& Kitsenko (blue solid). Top axis: conversion of $k_\perp$ into the angle of obliquity for $k_\parallel \lambda_{se}=0.05$. Right axis: power spectrum of the periodic \abrev{inner-foreshock} simulation (black dotted) for $k_\parallel \lambda_{se}=0.09$.}
    \label{figWhistlerGamma2}
\end{figure}

Fig.~\ref{figWhistlerGamma2} confirms that both the growth rate obtained from the hot-beams dispersion relation (green dashed) and the power spectrum of the parallel PBCS (black dotted) peak at the same value of $k_\perp$, although the parallel wave numbers of the maxima differ. For comparison, we also show an analytic approximation for gyroresonant whistler growth excited by a cold relativistic beam of electrons. As \citet{Zayed68} found, close to an anomalous gyroresonance this growth rate can be written as
\begin{equation}
\gamma_{\mathrm{ZK}} = \frac{\omres^{(\ell)}}2 \left(\frac{n_b}{n_0}\right)^{1/2} \frac{ (1-\kappa_\parallel) \left|\hat\lambda^2~\gamma_b^{-1}~(1+\hat\lambda^2) - \kappa_\parallel (1+\kappa_\parallel)\right|}{(1+\hat\lambda^2)~\kappa_\parallel^{3/2}},
\end{equation}
where $\kappa_\parallel = k_\parallel / |k|$ and $\hat\lambda = \lse^{-1}/|k|$. At least for the present case, this approximation for a cold beam overestimates the maximum growth rate by a factor of two, and severely underestimates the perpendicular wave number of the maximum.

\begin{figure*}
    \centering
    \includegraphics[width=.99\linewidth]{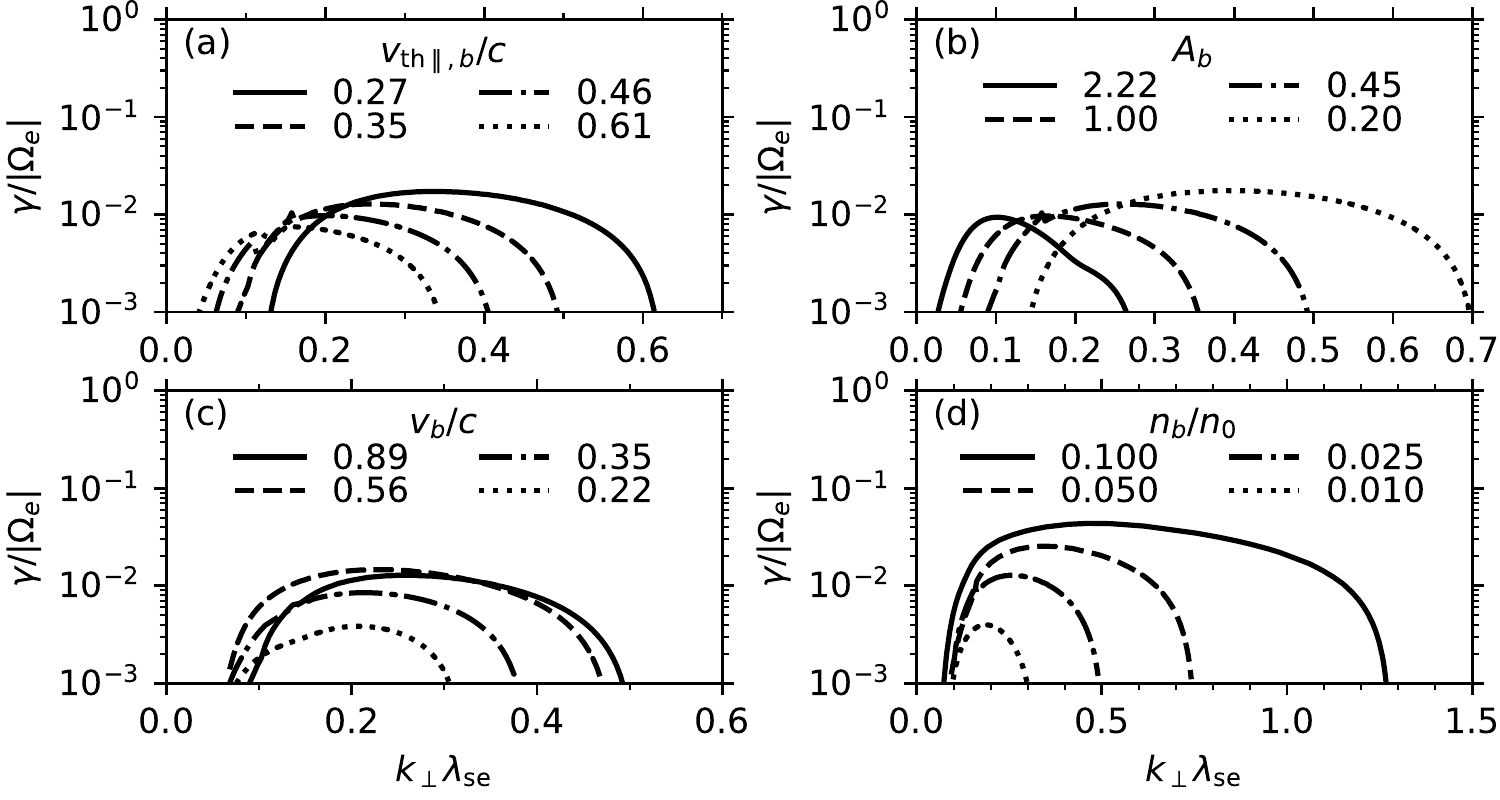}
    \caption{Growth rate of the oblique-whistler mode {as a function} of the perpendicular wave number, {evaluated at} the parallel wave number of the fastest-growing mode, i.e.\ of the second-order anomalous gyroresonance. {The growth rate is shown} for various values of the (a) thermal velocity $v_{\mathrm{th}\parallel,b}$, (b) temperature anisotropy $A_b$, (c) velocity $v_b$, and (d) density $n_b$ of the reflected electron beam. Unless given explicitly, {the parameter values} for the background ions and electrons are as stated in the text. {Those for the beam electrons are} $v_{\mathrm{th}\parallel,b}/c=0.27c, A_b=0.45, v_b/c=0.56, n_b/n_0=0.025$.}
    \label{figWhistlerGammaScan}
\end{figure*}

In general, the peak moves towards smaller perpendicular wave numbers as the temperature of the reflected beam decreases or as its temperature anisotropy $A_b$ increases. In addition to these trends, Fig.~\ref{figWhistlerGammaScan} shows that the perpendicular wave number of the fastest-growing mode depends only weakly on the parallel beam velocity, explaining why the linear calculation and the simulation agree so well in this aspect despite the different assumptions about the beam's phase-space distribution. The parallel wave number, on the other hand, is extremely sensitive to the parallel velocity distribution (cf.\ equation~(\ref{eqnOmegaRes})), which ensures that the excited waves are in resonance with the reflected electrons and scatter them efficiently. Decreasing the beam density of the reflected electrons leads to both a smaller obliquity and a smaller growth rate for the fastest-growing mode, which is why magnetic fluctuations are negligibly weak in the \abrev{outer foreshock}.

The properties of the background ions are, as one might expect, completely irrelevant for the oblique-whistler instability. The gyroresonant coupling that drives these waves occurs between the relatively cold background electrons and the hot beam of reflected electrons. For rightwards-travelling waves that are in gyroresonance with the background ions, one would expect the opposite helicity and polarisation (e.g.\citet{Weidl16}). Such a wave, in normal gyroresonance with the background ions because of their positive charge, would be necessary to initiate effective scattering of ions in the foreshock, as observed by \citet{Kumar21} in a one-dimensional PIC simulation. On the time scales of our two-dimensional shock setup, the magnetic fluctuations in the \abrev{inner foreshock} are due to oblique whistler waves. 

\section{Length of electrostatic and electromagnetic foreshocks} \label{sec:foreshocks-length}

\begin{figure}
    \centering
    \includegraphics[width=.99\linewidth]{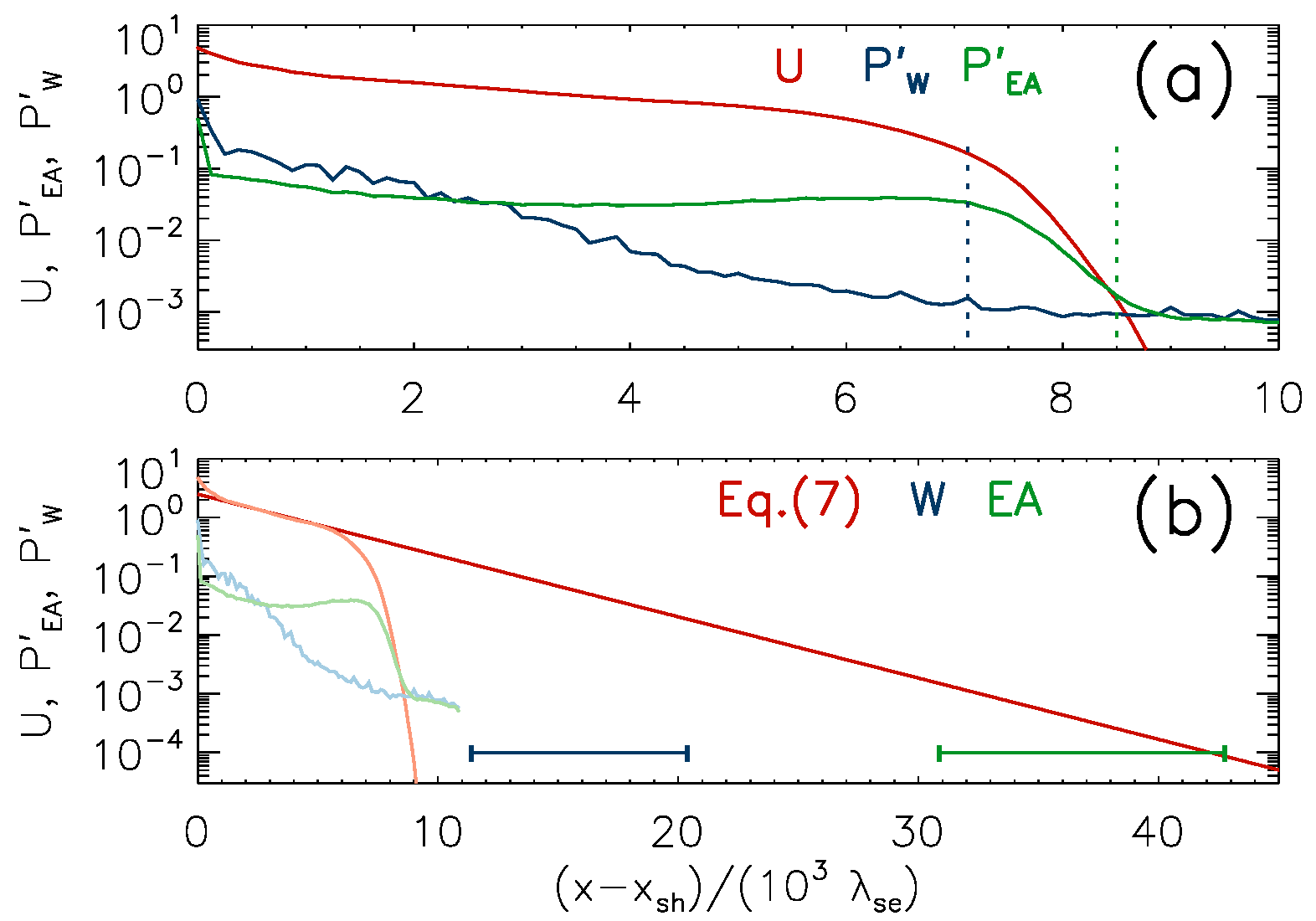}
    \caption{\ab{Panel (a): the normalised energy density of the shock-reflected electrons at the foreshock ($U = U_{e,r}/U_{e,sh}$, red curve); the modified power of electron-acoustic ($P'_{EA} = 10^2 P_{EA}$, green curve) and whistler waves ($P'_{w} = 10 \sqrt{P_{w}}$, blue curve); green and blue dotted lines shows where \mpon{the power of the electron-acoustic and the whistler waves is} twice the background noise level.
    Panel (b): the expected steady-state energy density distribution of the shock reflected electrons (Eq.~\ref{foreshock_distr}, red line); \abrev{green and blue} horizontal lines \mpon{mark the locations where the electron-acoustic (EA) and the} whistler (W) waves begin to be \mpon{emerge} in the steady state; the faded red, green, and blue curves \mpon{repeat the} lines from panel (a).} }
    \label{foreshock_plot}
\end{figure}

The reflected electrons propagate back upstream forming the extended foreshock. At the \mpo{end of the} simulation ($t = 50.4 \omci^{-1}$) the fastest electrons have reached the distance of about $9000 \lse$ from the shock position (Fig.~\ref{foreshock_plot}).
To identify where exactly electron-acoustic and whistler waves are excited in the \abrev{foreshock region}, we estimate the energy density of the corresponding modes \mpo{by summing the Fourier power in the $k$ space around the peak intensity, which can be read off} Fig.~\ref{figElectrostaticPBC}(a2) and \ref{figWhistlerProfile}(a2). \mpo{Specifically}, the power is 
\begin{equation}
P_{EA/W} = \sum_{k_\parallel} \sum_{k_\perp}  P(k_\parallel,k_\perp)\ , 
\end{equation}
where the \mpo{summation is performed in the wave-number range} $1<k_\parallel\lse<3$, $0<k_\perp\lse<3$ for electron-acoustic waves and $0.05<k_\parallel\lse<0.4$, $0.05<k_\perp\lse<0.4$ for whistler waves. Calculations are done over the \abrev{foreshock region}, $x>x_{sh}$, using a sliding window with the size of $x\times y = (125 \times 288)\lse$. Results are shown in Fig.~\ref{foreshock_plot} together with the bulk energy \mpo{density} of the shock-reflected electrons calculated in the upstream \ab{reference frame, $U_{e,r}$, and normalised by $U_{e,sh} = n_0\me c^2(\gamma_\mathrm{sh}-1)$, where $\gamma_\mathrm{sh} = 1/\sqrt{1-(\vsh/c)^2}$.}
Note that for a better visual representation we plot $P'_{EA} = 10^2 P_{EA}$ and $P'_{w} = 10 \sqrt{P_{w}}$.

The energy \mpo{density} of reflected electrons exponentially decreases with distance from the shock and can be described by the \pjm{empirical} function 
\begin{equation}
u(x) = 2.5 (1+2.4 \cdot 10^{-4})^{-(x-x_{sh})/\lse},
\label{foreshock_distr}
\end{equation}
\pjm{which is shown by} the red line in Fig.~\ref{foreshock_plot}(b)). Note that the numerical coefficients remain stable over at least the last $10\omci^{-1}$ of the shock simulation, which permits \mpo{a reliable} extrapolation to later stages.

The growth rate of the electron-acoustic waves is high, and therefore their front propagates together with the fastest electrons. The power of the electron-acoustic waves is twice \mpo{the} background noise level at $(x-x_{sh}) \approx 8500 \lse$ where the \mpo{energy density of} reflected electron reaches the value of $U \approx 1.5\cdot10^{-3}$ (the green dotted line in Fig.~\ref{foreshock_plot}(a)). Whistler waves grow much \mpo{more slowly} and require stronger currents to be driven. They become visible only when \mpo{at $(x-x_{sh}) \approx 7100 \lse$ the energy density} of reflected electrons reaches $U \approx 0.16$ (the blue dotted line in Fig.~\ref{foreshock_plot}(a)).

\mpo{The simulation time limits the distance from the shock that the shock-reflected electrons can reach}. Knowing that \mpo{the electrons have a stable distribution we may extrapolate} it further upstream. In this way we can mimic the steady state of the foreshock and can find where the electron-acoustic and the whistler waves will be present if the simulation \mpo{were} continued further. 

The threshold conditions at which electron-acoustic and whistler waves are excited can be defined in two ways. Either waves start growing when $U$ reaches a critical value, as discussed above, or we need to account for the \mpo{integrated energy density of shock-reflected electrons up to the point where wave growth} becomes detectable. If the second definition applies, then waves can be triggered further upstream when the shock reaches its steady state.

Using the properties of reflected electrons at $t=50.4\omci^{-1}$ \ab{and Eq.~\ref{foreshock_distr}}, we \mpo{can estimate the regions where wave growth commences in} the steady state. They are marked by green \ab{(electron acoustic waves, EA) and blue (whistler waves, W)} horizontal lines in Fig.~\ref{foreshock_plot}(b). One would require to run the shock simulation up to $125\omci^{-1}$ to fully cover the whistler region and to about $270\omci^{-1}$ to cover the entire electron foreshock and reach the steady-state stage of the shock evolution.

\section{Conclusions} \label{sec:summary}

\ab{
In this paper we report results of 2D PIC simulations of an oblique high-Mach-number shock with $\thbn = 60^\circ$ and $M_A=30$. The simulation parameters are chosen to study the formation of the electron foreshock at conditions close to those in SNRs \mpon{and for comparison with} 1D PIC simulations with $M_A=63$ \cite{Xu2020}. Using linear theory and PIC simulations with periodic boundary conditions, we have identified the dominant waves produced by the shock-reflected electrons in the foreshock region.}

\ab{Our results can be summarised as follows:}

\begin{itemize}
    \item \ab{In the shock simulation a fraction of incident electrons are reflected back upstream with velocities \pjm{significantly} exceeding the shock speed. These electrons travel along the upstream magnetic field and generate electrostatic waves \abrev{far from the shock} and electromagnetic waves closer to the shock, forming the so-called electron foreshock.}
    
    \item The electrostatic waves \abrev{in the outer foreshock} are generated by the electron-acoustic instability. After modifying its linear dispersion relation to account for a Gaussian momentum distribution, we find excellent agreement between linear theory and the parallel PBCS. Although solving this modified dispersion relation is difficult for oblique propagation directions, we attribute the \abrev{outer-foreshock} waves that we observe in the oblique PBCS to the electron-acoustic instability as well.
    
    \item The electromagnetic waves in the \abrev{inner foreshock} have the same perpendicular wave number and growth rate as linear theory predicts for the fastest-growing mode of the oblique-whistler instability. This electron-electron mode can explain our PBCS observations very well even though we have used a bi-Maxwellian velocity distribution in the dispersion relation and a bi-Gaussian momentum distribution in the simulations. However, the resulting difference in the parallel velocity distribution shifts the spectral peak from the second-order gyroresonance (according to the linear model) to a larger parallel wave number close to third-order gyroresonance (in the parallel PBCS).
    
   \item Over the total run time of the simulation (more than $50~\omci^{-1}$) we do not see any indication that the reflected electrons are exciting collective motion of the counterpropagating ions in the foreshock: although the ions are being stochastically heated in the \abrev{foreshock} turbulence, where $v_{\mathrm{th}\parallel,i}$ increases by about 50\% overall, we do not detect ion/electron modes like the Buneman instability. Both the electron-acoustic and the oblique-whistler instability are predominantly driven by the relative drift between background electrons and reflected electrons. After this relative drift has been reduced through wave-particle interactions, the electron distribution is too close to isotropic to excite an anistropy-driven mode like the whistler heat-flux instability.

    \item \ab{Electron-acoustic and whistler waves interact with electrons in the foreshock and maintain a steady-state density profile which propagates upstream in front of the shock. The energy of reflected electrons \mpon{scales} exponentially with distance from the shock and can be approximated as $U_{e,r}/U_{e,0} = 2.5 (1+2.4 \cdot 10^{-4})^{-(x-x_{sh})}$, where the distance $x-x_{sh}$ is given in $\lse$. For the \mpon{shock parameters} discussed in this paper, the shock simulations should be continued up to $270\omci^{-1}$ to cover the entire electron foreshock and reach \mpon{its} steady state.}
    
\end{itemize}

\ab{\mpon{The instabilities discussed in this paper operate on} the electron scale. \mpon{Considering that} the typical scale of a shock is the ion Larmor radius, the time needed to reach the steady state depends on the upstream plasma parameters. Assuming that the shock-reflected electrons move from the shock with some speed proportional to the shock velocity, and the foreshock length is roughly the same as shown above, we can deduce that \mpon{the time needed to establish the steady state,} $t \propto \lse/\vsh \propto \ma^{-1}\sqrt{\me/\mi}$, \mpon{is a smaller multiple of the ion gyrotime, $\omci^{-1}$, if the real proton-to-electron mass ratio is applied or shocks of higher Alfvenic Mach number are considered.}}

The waves generated in the electron foreshock are capable of scattering electrons proposing new channels of electron heating, acceleration and even injection into DSA, which require special attention in the future.
\abrev{We have discussed the results of the 2D3V simulation, but the real world is 3D throughout. When performed in 3D the shock simulation should reveal realistic densities and momentum distributions of electrons responsible for formation of the electron foreshock. This will define the length of the foreshock and intensity of electron-acoustic and whistler waves, and it will affect the particle dynamics in the foreshock. However, the general structure of the foreshock should be as described here, on account of the large difference in the growth rate of the instabilities discussed here.}

\begin{acknowledgments}
M.P. acknowledges support by DFG through grant PO 1508/10-1.
The numerical experiments were done with HLRN supercomputer at North-German Supercomputing Alliance under the project bbp00033.
\end{acknowledgments}

\appendix

\section{modified electron-acoustic instability}
\label{appEAIp}

Our PBC simulations initialize the distribution of each particle species as bi-Gaussian in momentum. The more common assumption that the distribution is bi-Gaussian in velocity (which our {\scshape Whamp} calculations employ) results in unphysical contributions of velocities larger than the speed of light because of the large relativistic drift of the reflected electrons. At least for the electron-acoustic instability in the weakly unstable limit, however, it is straightforward to derive a growth rate for a Gaussian distribution of $p_\parallel$.

We start with the dielectric permittivity $\epsilon(\omega,k)$ for longitudinal electrostatic waves in a uniform plasma composed of cold background electrons and an additional electron beam. After the usual Fourier--Laplace transform in space and time, we obtain for $k\in\mathbb R$ and $\omega\in \mathbb C$
\begin{equation}
    \epsilon(\omega,k) = 1 - \frac{\ompe^2}{\omega^2} + \chi_b(\omega,k).
    \label{eqnEpsilon}
\end{equation}
Here we assume that all ion terms can be neglected for the frequencies which are of interest ($|\omega|\gg\ompi$). The first two terms on the right-hand side form the dielectric permittivity of the cold electron background; the susceptibility $\chi_b$ of the reflected-electron beam reads
\begin{equation}
    \chi_b(\omega,k) = -\frac{\ompb^2}{k^2} \int_{-\infty}^{+\infty} \frac{\mathrm{d}u}{\gamma_u \frac{\omega}{k} - u} f_b'(u),
\end{equation}
where $\ompe$ and $\ompb=(n_b/n_e)^{1/2}\ompe$ are the plasma frequencies of the two electron species (without relativistic correction), $u=p_\parallel/m$ is the normalized parallel momentum, and $\gamma_u=(1+u^2)^{1/2}$ denotes the corresponding Lorentz factor. For a Gaussian distribution in $u$, we write
\begin{equation}
    f_b(u) = (2\pi u_{\mathrm{th}}^2)^{-1/2} \exp\left[-\frac12 \left(\frac{u-u_0}{u_{\mathrm{th}}}\right)^2\right].
\end{equation}
After following van~Kampen's treatment of Landau damping to solve the integral through contour integration, analytic continuation of the integrand, and application of the Sochocki--Plemelj theorem, we obtain for the weakly unstable limit $\Im\omega\to0$
\begin{equation}
    \chi_b(\omega,k) = -\frac{\ompb^2}{k^2} \left[\mathcal P \int_{-\infty}^{+\infty} \frac{\mathrm{d}u}{\gamma_u \frac{\omega}{k} - u} f_b'(u) - 2\pi \II f_b'(u_{\omega/k})\right],
\end{equation}
where the pole $u_{\omega/k}=(1-\omega^2/c^2k^2)^{-1/2}\omega/k$ is the normalized momentum for which an electron is in Landau resonance with a wave propagating with the phase speed $\omega/k$. This is where the Gaussian \emph{momentum} distribution leads to a critical difference compared to the standard electron-acoustic instability with a Gaussian \emph{velocity} distribution: only in the former case is the relativistic Lorentz factor $(1-\omega^2/c^2k^2)^{1/2}$ included in the definition of the pole $u_{\omega/k}$.

For a sufficiently tenuous beam ($\ompb\ll\ompe$), we can ignore the beam contribution as we use (\ref{eqnEpsilon}) to solve $\epsilon(\omega,k)\equiv0$ for the angular frequency $\varpi=\Re\omega$. The result, a simple plasma oscillation with $\varpi\equiv\ompe$, determines the real frequency that we use in the next step.

Of course, we must include the beam contribution in order to compute the growth rate of the instability. Following Landau again, we differentiate $\Im\epsilon=0$ with respect to $\omega=\varpi+\II\gamma$ and find in the weakly unstable limit $\gamma\to0$:
\begin{equation}
    \gamma(k) = -\left.\Im\epsilon(\ompe,k) \middle/ \frac{\partial\Re\epsilon}{\partial\omega}(\ompe,k)\right. .
\end{equation}
We can now neglect the principal value $\mathcal P$ in the numerator because it is real and the beam contribution in $\Re\epsilon$ because the background electrons are much more important for the angular frequency. The final result for the growth rate of the Gaussian-momentum electron-acoustic instability is given by equation~(\ref{eqnRelElAcoustic}).

In Fig.~\ref{figElectrostaticGamma}, we also plot the (in $n_b/n_e$) first-order solution of the angular frequency $\varpi$. It is obtained from expanding the resonance denominator of the principal value to second order around the peak of the Gaussian $u_0$ and numerically solving $\Re\epsilon=0$ for $\varpi$, where
\begin{equation}
    \Re\epsilon(\varpi,k)=1-\frac{\ompe^2}{\varpi^2}-\frac{\ompb^2}{k^2}\ \frac{1+3\left(\frac{u_{\mathrm{th}}}{u_{\varpi/k}-u_0}\right)^2}{(u_{\varpi/k}-u_0)^2}.
    \label{eqnRealFreqEAI}
\end{equation}

\bibliography{aipsamp}

\end{document}